\documentclass[
reprint,
superscriptaddress,
amsmath,amssymb, 
aps,
]{revtex4-2}

\usepackage{graphicx}
\usepackage{amsmath}
\usepackage{natbib}
\usepackage{amssymb}
\usepackage{braket}
\usepackage{hyperref}
\usepackage{array}
\usepackage{float}
\usepackage[usenames,dvipsnames]{color}
\usepackage{graphicx}
\usepackage{subfigure}
\usepackage{float}
\usepackage{verbatim}
\usepackage{xcolor}
\usepackage{cleveref}
\usepackage{pdfsync}
\usepackage[version=4]{mhchem}
\usepackage{chemarr}
\usepackage{mathtools}
\usepackage[export]{adjustbox}
\begin{document}

\begin{@twocolumnfalse}
\title{Kinetic theory of emulsions with matter supply}

\author{Jacqueline Janssen}
\affiliation{Laboratoire de Physique de l’{\'E}cole Normale Sup{\'e}rieure, ENS, Universit{\'e} PSL, CNRS, Sorbonne Universit{\'e},
Universit{\'e} Paris Cit{\'e}, 75005 Paris, France
}
\affiliation{Max Planck Institute for the Physics of Complex Systems,
N{\"o}thnitzer Stra{\ss}e~38, 01187 Dresden, Germany
}

\author{Frank J\"ulicher}
\affiliation{Max Planck Institute for the Physics of Complex Systems,
N{\"o}thnitzer Stra{\ss}e~38, 01187 Dresden, Germany
}
\affiliation{Center for Systems Biology Dresden, Pfotenhauerstra{\ss}e~108, 01307 Dresden, Germany
}
\affiliation{Cluster of Excellence Physics of Life, TU Dresden, 01062 Dresden, Germany
}

\author{Christoph A. Weber}
\affiliation{Faculty of Mathematics, Natural Sciences, and Materials Engineering: Institute of Physics, University of Augsburg, Universit{\"a}tsstra{\ss}e~1, 86159 Augsburg, Germany
}

\date{\today}

\begin{abstract}
In this work, we propose a theory for the kinetics of emulsions in which a continuous supply of matter feeds droplet growth. 
We consider cases where growth is either limited by bulk diffusion or the transport through the droplets' interfaces. 
Our theory extends the Lifshitz-Slyozov-Wagner (LSW) theory by two types of matter supply, where either the supersaturation is maintained or the supply rate is constant.
In emulsions with maintained supersaturation,
we find a decoupling of droplets at all times, with the droplet size distribution narrowing in the diffusion-limited regime and a drifting distribution of a fixed shape in the interface-resistance-limited case.
In emulsions with a constant matter supply, there is a transition between narrowing and broadening in the diffusion-limited regime, and the distribution is non-universal.
For the interface-resistance-limited regime, there is no transition to narrowing, and we find a universal law governing coarsening kinetics that is valid for any constant matter supply. 
The average radius evolves according to a power law that is independent of the matter supply, and we find a closed-form expression for the droplet size distribution function.
Our theory is relevant to biological systems, such as biomolecular condensates in living cells, since droplet material is not conserved and the growth of small droplets is proposed to be interface-resistance-limited. 
\end{abstract}
\maketitle

\end{@twocolumnfalse}

\section{Introduction}

When a liquid mixture of immiscible components undergoes phase separation, many small droplets are nucleated~\cite{bray_theory_1994, barrat2003basic, livi2017nonequilibrium}. 
This emulsion coarsens with the average droplet size increasing over time. 
Such coarsening typically involves the growth of larger droplets at the expense of smaller ones, which shrink and eventually dissolve, a phenomenon commonly referred to as Ostwald ripening. On long-time scales,
a finite system settles at thermodynamic equilibrium corresponding to a single droplet. 

Coarsening kinetics is relevant across various length-scales and occurs in very different systems, ranging from cloud formation and initiation of precipitation~\cite{pruppacher_microphysics_1998}, over annealing of polymer blends~\cite{hamley2008introduction}, exsolution in feldspar minerals~\cite{brown_exsolution_1984}, to the coarsening of biomolecular condensates such as P granules or stress granules in living cells~\cite{hyman_beyond_2012, berry_physical_2018,banani_biomolecular_2017}. Interestingly, such biomolecular condensates behave like liquid droplets that can fuse and grow by diffusive influx through droplet interfaces.
For example, nuclear bodies dominantly coarsen via
fusion, and the transport through the interface is the rate-limiting process~\cite{stephaniecliff}.  
In general, ripening inside a living cell is expected to differ from classical ripening in passive systems that tend toward thermodynamic equilibrium, 
since the cellular environment is maintained away from equilibrium.
Among others, a crucial difference to passive systems involves the synthesis of droplet components and, thereby, the supply of droplet matter.

\begin{figure}[b]
    \centering
    \includegraphics[width=1\linewidth]{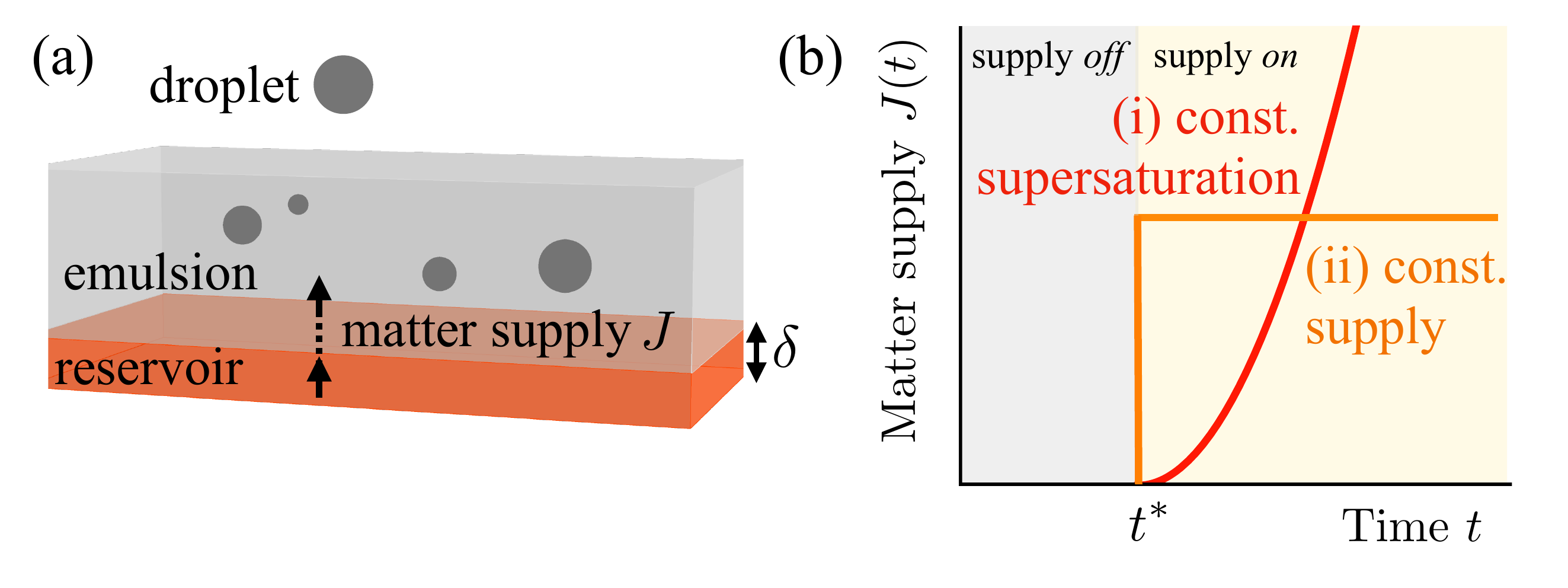}
    \caption{\textbf{Emulsion supplied by droplet matter.} 
    (a) Illustration of an emulsion coupled to a reservoir (red) supplying the system with matter, leading to growing droplets.
    (b) 
    The matter supply $J$ is switched on at time $t^*$. 
    Supply cases: 
    (i) To keep the supersaturation constant,
    the matter supply $J\propto t^2$ increases since droplets grow in time.
    (ii) Constant matter supply (orange).
    }
    \label{fig1}
\end{figure}

In the physics literature, the kinetics of coarsening is distinguished by total droplet matter being either conserved or non-conserved~\cite{bray_theory_1994, livi2017nonequilibrium}. 
In a non-conserved system, droplet matter can be locally created or destroyed. 
The classical example is the Ising model, where the magnetization is not conserved as spin orientations can flip. These spin flips can move the interface and drive coarsening, favoring domains with smaller surface-to-volume ratios whose average size grows $\langle R \rangle \propto t^{1/2}$. 
In a conserved system, total droplet matter can only be redistributed in space, for example, by diffusion. 
Coarsening in conserved emulsions where molecular components diffuse is governed by the Lifshitz-Slyozov and Wagner theory (LSW)~\cite{lifshitz_kinetics_1961,wagner_theorie_1961}. 
The classical LSW theory describes Ostwald ripening~\cite{ostwald_studien_1897,ostwald_lehrbuch_1896}, which corresponds to the diffusion-limited coarsening of an emulsion with a broadening droplet size distribution.
The average radius follows a power-law, $\langle R \rangle \propto t^{1/3}$,
by which the distribution can be rescaled, yielding a universal shape that is independent of physical parameters and the initial condition. 

Wagner also formulated the ripening theory of coarsening in the interface-resistance-limited regime~\cite{wagner_theorie_1961}, which arises when the transfer of mass across the interface is slower than the diffusion of material toward the interface. This regime is relevant at the early stage of coarsening when droplets are small, such that the transfer of material across the interface is slower than the relaxation of the concentration gradients. 
The key result of Wagner's theory is a universal and broadening droplet size distribution with an average droplet size following a power-law $\langle R \rangle \propto t^{1/2}$.
The time scales associated with the transport across the interface are governed by interfacial resistance.
Recently, interfacial resistance has been suggested to be relevant in biomolecular condensate dynamics~\cite{folkmann_regulation_2021, PhysRevResearch.3.043081, zhang_exchange_2024, hubatsch_transport_2024} and 
coacervate systems~\cite{munchow_protein_2008, hahn_size-dependent_2011,hahn_electrophoretic_2011}.

When emulsions are maintained away from equilibrium by the supply of energy or matter, they are called active~\cite{weber_physics_2019, julicher2024droplet, zwicker2024chemically, cates2025active}.
Their coarsening can differ from that of passive emulsions, as their dynamics are governed by both conserved and non-conserved fields~\cite{bauermann2025dropletsize}. 
Mixtures with chemical reactions can realize such active emulsions maintained away from equilibrium~\cite{weber_physics_2019}, giving rise to novel dynamic behaviors that include the spontaneous formation of stable liquid shells~\cite{bergmann_liquid_2023, shells-theory2023}, droplet division~\cite{zwicker_growth_2017}, accelerated ripening~\cite{tena2021accelerated}, or suppression of Ostwald ripening~\cite{zwicker_suppression_2015}, as well as bubbly phase separation~\cite{cesare2018}.
More recently, dynamic theories for emulsion dynamics have been proposed, leading to ripening arrest~\cite{chiu-fan-sizecontrol2018} and reversed ripening~\cite{cesare2024, bauermann2025dropletsize}.
Emulsions maintained away from equilibrium by constant matter supply in the diffusion-limited regime were shown to give a narrowing and non-universal droplet size distribution, where the prefactor and the scaling exponent of the moments depend on the driving strength~\cite{vollmer_ripening_2014, clark_focusing_2011}.
However, it remains elusive how 
emulsions with matter supply differ between diffusion- and interface-resistance-limited transport, as well as how different matter supplies relevant for biological systems and chemical systems affect the emulsion dynamics.

In this work, we present a theoretical framework which extends the LSW theory~\cite{lifshitz_kinetics_1961,wagner_theorie_1961} to emulsions subject to two types of matter supply: (i) constant supersaturation and (ii) constant matter supply (Fig.~\ref{fig1}(a,b)).
In this framework, we account for the interplay between diffusive fluxes from the bulk and molecular transport through the interface, which both drive the growth and shrinkage of droplets. 
Our theory can explain the time evolution of droplet size distributions for each type of matter supply and unravel the physical mechanisms underlying the universality of coarsening in emulsions with matter supply.

\section{Theory for emulsion kinetics with matter supply}
\label{sec:theory}

\subsection{Emulsion dynamics composed of $N$ droplets}
\label{eq:discrete_model}

The emulsion consists of $N(t)$ spherical droplets of radii $R_i$ ($i=1,...,N$)
in a volume $V_\text{sys}$.
The total droplet phase volume is thus given by $V_\text{d}(t) = \sum_{i = 1}^{N(t)} (4\pi/3) R_i(t)^3$. We consider droplets as dilute in the system, implying that the total droplet volume is small compared to the system size, $V_\text{d}(t) \ll V_\text{sys}$. 
Concentrations inside droplets are considered to be homogeneous and constant, given by the equilibrium value $c^\text{in} = c^\text{in,(0)}$. 

Droplets are coupled via a coarse-grained  homogeneous background concentration $\bar{c}(t)$
that exceeds the dilute phase's equilibrium concentration $c^{(0)}$ (supersaturated case) ensuring thermodynamic stability of droplets.
The coarse-grained background concentration of the dilute phase evolves according to
\begin{equation}
       \frac{\text{d}}{\text{d} t}  \bar{c}(t) = -\frac{c^\text{in,(0)}}{V_\text{sys}} \frac{\text{d}}{\text{d}t} \sum_{i=1}^{N(t)} \frac{4 \pi}{3} R_i(t)^3 + J(t)\,,
    \label{eq:cons-law-discrete}
\end{equation}
where the first term describes the gain/loss contributions when droplets shrink/grow, and $J(t)$ is the external matter supply density (see Appendix~\ref{app:calc-barc}). 

We consider two cases of supply:
\begin{enumerate}
\label{item:matter_supply}
\item[(i)] The matter supply density $J(t) = c^\text{in,(0)}V_\text{sys}^{-1} \, \text{d} {V}_\text{d}(t)/\text{d}t$ with $\text{d}\bar c/\text{d}t=0$, that fixes the coarse-grained concentration $\bar{c}$, maintaining supersaturation
\begin{equation}
\label{eq:const_supersaturation}
\varepsilon=\frac{\bar{c}}{c^{(0)}}-1 = \text{const} \, . 
\end{equation}

\item[(ii)]  Constant matter supply, \begin{equation}
\label{eq:const_flux}
J = \text{const}\, ,
\end{equation}
with $J = 0$ corresponding to the case of a passive emulsion without matter supply. 
\end{enumerate}

\begin{figure*}[htb]
    \centering
    \includegraphics[width=1\linewidth]{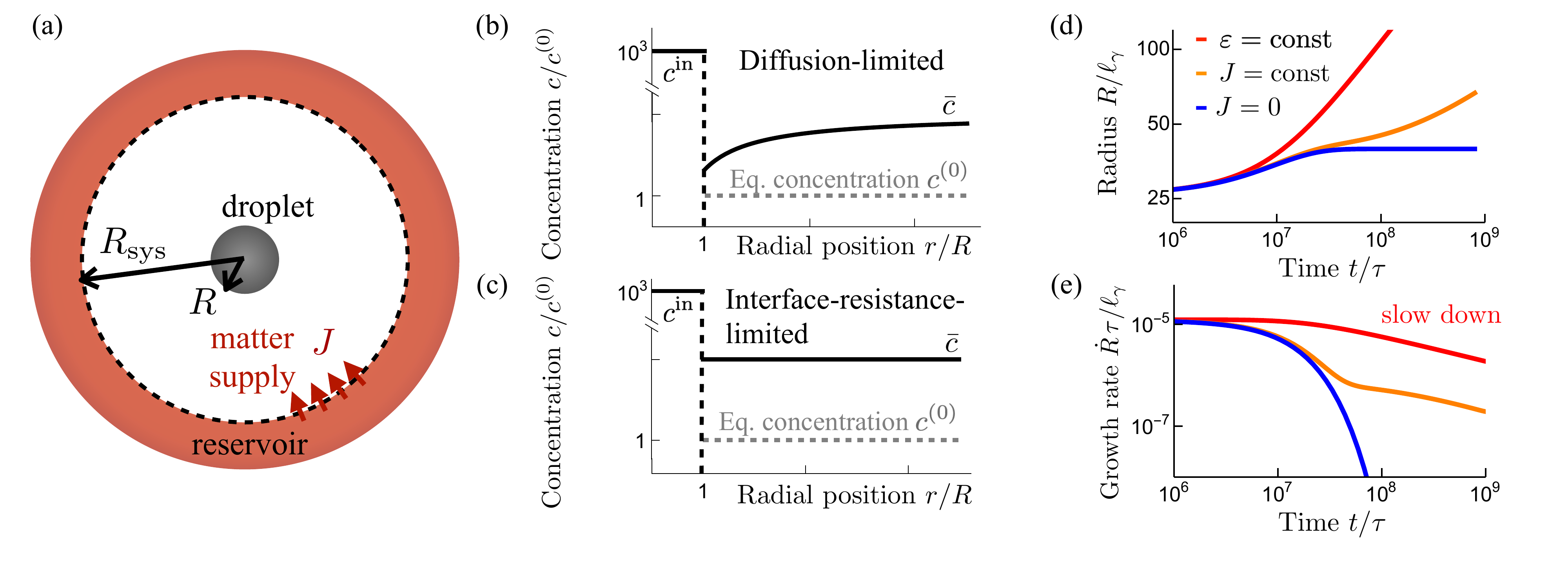}
    \caption{\textbf{Single droplet in a system coupled to a material reservoir.} 
    (a) Illustration of a droplet of radius $R$ in a system of size $R_\text{sys} \gg R$. Far away from the droplet interface, at the system's boundary, the matter is supplied from the reservoir. Due to fast diffusion, we assume no spatial gradients of the droplet material concentration at the system's boundary. (b) Illustration of the concentration profile in the diffusion-limited regime. The concentration is constant inside the droplet, while outside it relaxes to $\bar{c}$ with a spatial dependence $\propto R/r$. (c) Illustration of the concentration profile in the interface-resistance-limited regime. Due to fast diffusion, the concentration outside relaxes instantaneously to $\bar{c}$. (d) Growth of the droplet in the diffusion-limited regime for the three supply cases. In the presence of a matter supply, the radius grows indefinitely, while in the passive case, the growth ceases close to equilibrium. $\tau = \ell_\gamma^2/D^\text{out}$ in the log-log plot; $\ell_\gamma$ denotes the capillary length.  
    (e) Growth speed of the droplet radius $R$ as a function of time in the log-log plot. In the case of constant supersaturation, there is a regime of accelerated growth that decreases and follows the $R^{-1}$ law at late times. The same slowdown is valid for a constant matter supply, and it starts when the droplet is large enough such that the $R^{-1}$ term in the growth law dominates the $R^{-2}$ term. For the passive case, the growth speed decreases as $R^{-2}$. Results shown in panels (d,e) were obtained solving Eqs.~\eqref{eq:growth1-conc1}, for the diffusion-limited regime with the parameters given in Table~\ref{tab-param1}.}
    \label{fig2-diff-lim}
\end{figure*}

For an emulsion with a constant equilibrium concentration inside $c^{\text{in},(0)} \gg c^{(0)}$ (strong phase separation), the interface speed of droplet $i$ reads 
\begin{equation}
    \frac{\text{d}}{\text{d} t} R_i(t) = -\frac{\left.j^\text{out}\right|_{R_i}}{c^\text{in,(0)}} \,,
    \label{eq-drdt}
\end{equation}
where $\left.j^\text{out}\right.|_{R_i}$ is the normal flux evaluated outside at the interface in the lab frame. This flux $\left.j^\text{out}\right|_{R_i}$ can be determined in the near-field using a quasi-static approximation, where transients in the diffusion equation vanish. In this approximation, the concentration field satisfies a Laplace equation from which the near-field concentration can be obtained. The resulting concentration
at the interface 
\begin{equation}
    c(R_i) = \frac{D^\text{out} \bar{c}(t) + k R_i c^\text{eq}(R_i)}{D^\text{out}+ kR_i}\,,
    \label{eq-c-boundary}
\end{equation}
is characterized by the diffusion constant of the dilute phase $D^\text{out}$ and the interfacial conductivity  
$k$ describing an effect referred to as interfacial resistance~\cite{folkmann_regulation_2021, PhysRevResearch.3.043081, zhang_exchange_2024, hubatsch_transport_2024, munchow_protein_2008, hahn_size-dependent_2011,hahn_electrophoretic_2011}.
The flux at the interface  
\begin{equation}
    \left.j^\text{out}\right|_{R_i} = k \Big({c}^{\text{eq}}(R_i)- c(R_i)\Big)\,,
    \label{eq:flux}
\end{equation}
is governed by interfacial conductivity  
$k$ in response to the thermodynamic driving force, which is the deviation of concentration at the interface from its equilibrium value (see Appendix~\ref{app-intlim-dRdt}).
For a finite droplet, this equilibrium concentration is given by the 
the Gibbs-Thomson relationship
\begin{equation}
\label{eq:ceq_finite}
c^\text{eq}(R_i) = c^{(0)}\left(1+\frac{\ell_\gamma}{R_i}\right) \, .  
\end{equation}
It describes how the equilibrium concentration 
$c^{(0)}$ of the dilute phase for a flat interface (thermodynamic limit) is affected by droplet size $R_i$, with
the capillary length $\ell_\gamma = 2\gamma/(k_\text{B} T c^\text{in,(0)})$, and $\gamma$ denoting the surface tension. 
Eqs.~\eqref{eq-c-boundary} and \eqref{eq:flux} imply that the non-dimensional quantity 
\begin{equation}
\label{eq:beta}
   \beta(R_i) = \frac{k R_i}{D^\text{out}} \, ,
\end{equation}
characterizes the transition between diffusion-limited transport ($\beta \gg 1$) and interface-resistance-limited transport ($\beta \ll 1$).
We note that for large radii $R_i$, $\beta \gg 1$, and thus droplet growth is always diffusion-limited when droplets are big enough.
Similarly, we introduce a non-dimensional quantity 
\begin{equation}
\label{eq:beta-ensemble}
   \bar{\beta}(\langle R \rangle) = \frac{k \langle R \rangle}{D^\text{out}} \, ,
\end{equation}
for an ensemble of droplets with a mean radius $\langle R \rangle= \sum_i^N R_i/N$. This quantity distinguishes diffusion-limited transport ($\bar{\beta} \gg 1$) and interface-resistance-limited transport ($\bar{\beta} \ll 1$) for an ensemble of many droplets.

The growth of droplet $i$ accounting for both transport by  diffusion and interface-resistance is given by
\begin{equation}
     \frac{\text{d}}{\text{d} t} R_i(t) = \frac{k}{1 + \beta\left(R_i\left(t\right)\right)} \frac{1}{c^\text{in,(0)}}\Big(\bar{c}(t) -c^\text{eq}\left(R_i\left(t\right)\right)\Big)\,.
    \label{eq-drdt0}
\end{equation}
For diffusion-limited transport ($\beta \gg 1$), we find~\cite{lifshitz_kinetics_1961,wagner_theorie_1961}
\begin{equation}
     \frac{\text{d}}{\text{d} t} R_i(t) = \frac{D^\text{out}}{R_i\left(t\right)} \frac{1}{c^\text{in,(0)}}\Big(\bar{c}(t) -c^\text{eq}\left(R_i\left(t\right)\right)\Big)\,,
     \label{eq:drdt-difflim}
\end{equation}
while for interface-resistance-limited transport ($\beta \ll 1$) \cite{wagner_theorie_1961},
\begin{equation}
     \frac{\text{d}}{\text{d} t} R_i(t) =   \frac{k}{c^\text{in,(0)}}\Big(\bar{c}(t)-c^\text{eq}\left(R_i\left(t\right)\right)\Big)\,.
\end{equation}
In summary, the kinetics of emulsions with matter supply is governed by the dynamics of the coarse-grained concentration (Eq.~\eqref{eq:cons-law-discrete}), coupled to the dynamic equations for each droplet radius (Eqs.~\eqref{eq-drdt0}).
We use these equations to study the dynamics of a single droplet (Sect.~\ref{sec-one}) and two droplets (Sect.~\ref{sec-two}) in the presence of matter supply. 
In Sect.~\ref{sec-emulsion}, we consider emulsions in the thermodynamic limit ($N\to \infty$, $N/V_\text{sys}$ fixed) and derive the corresponding continuum theory to obtain the dynamic droplet size distribution with matter supply. 

\section{Single droplet dynamics}
\label{sec-one}

The case of a single droplet ($N=1$) is depicted in Fig.~\ref{fig2-diff-lim}(a). 
The dynamic Eqs.~\eqref{eq:cons-law-discrete} and \eqref{eq-drdt0} take for $N=1$ the following form:
\begin{subequations}
\begin{align}
      \frac{\text{d}}{\text{d} t}  \bar{c}(t) &= -\frac{{4 \pi} c^\text{in,(0)}}{V_\text{sys}} R(t)^2 \, \frac{\text{d}R\left(t\right)}{\text{d}t} + J(t)\,, 
    \label{eq:conc1}\\
     \frac{\text{d}}{\text{d} t} R(t) &= \frac{k}{1 + \beta\left(R\left(t\right)\right)} \frac{1}{c^\text{in,(0)}}\Big(\bar{c}(t) -c^\text{eq}\left(R\left(t\right)\right)\Big)\,,
     \label{eq:growth1}
\end{align}
\label{eq:growth1-conc1}
\end{subequations}
with the concentration
$c^\text{eq}(R)$ at the droplet  interface $R$ given by Eq.~\eqref{eq:ceq_finite}
and the non-dimensional quantity $\beta$ defined in Eq.~\eqref{eq:beta}.

Figs.~\ref{fig2-diff-lim}(b) and (c) show the difference in the concentration profile in the two regimes: (b) Diffusion-limited growth ($\beta \to \infty$) and (c) interface-resistance-limited growth  ($\beta \rightarrow 0$), respectively. 
In the diffusion-limited regime (Fig.~\ref{fig2-diff-lim}(b)), the concentration profile has a radial dependence.
For the case of droplet growth, the concentration increases from the interface and approaches the coarse-grained dilute phase concentration $\bar{c}$ far away from the interface. In contrast, in the interface-resistance-limited regime (Fig.~\ref{fig2-diff-lim}(c)), 
the concentration profile is approximately flat and equal to $\bar{c}$ since 
diffusion is fast compared to the rate-limiting transport through the interface. 
These features are related to the near-field and arise from the quasi-static assumption and the boundary condition far from the interface, $\bar{c}$. 
Since matter supply exclusively affects the coarse-grained dilute phase concentration $\bar{c}$, the near-field concentration profiles depicted in Figs.~\ref{fig2-diff-lim}(b,c) are independent of the type of matter supply (see Eqs.~\eqref{eq:const_supersaturation} and \eqref{eq:const_flux}).

We now discuss single droplet growth in the diffusion-limited regime (see Fig.~\ref{fig2-diff-lim}(d,e)). 
The results for the interface-resistance-limited regime are depicted in  Figs.~\ref{int-lim1}(a,b) in the Appendix.
We find that the presence of matter supply strongly alters the droplet growth kinetics of a single droplet relative to the passive case.   
In the passive case, the droplet can only grow by picking up the droplet material that exceeds the Gibbs-Thomson concentration $c^\text{eq}(R)$, leading to a vanishing droplet growth 
and a stationary droplet radius on long times when the excess concentration is depleted (blue line in Fig.~\ref{fig2-diff-lim}(d,e)). 

Supplying matter by maintaining the supersaturation constant (Eq.~\eqref{eq:const_supersaturation}), or supplying matter with a constant flux density $J$ (Eq.~\eqref{eq:const_flux}), gives rise to a fast and indefinite growth of the droplet (red and orange line in Fig.~\ref{fig2-diff-lim}(d,e), respectively) compared to the passive system. 
The growth rate remains finite and decreases slowly. This decrease arises from the $R^{-1}$ term in Eq.~\eqref{eq:growth1} that dominates the growth law at late times. 
In other words, for diffusion-limited transport, an infinitely large droplet will stop growing even when matter is supplied by maintaining a fixed supersaturation. 

The growth of a single droplet in the interface-resistance-limited regime is different (see Appendix Figs.~\ref{int-lim1}(a,b)). 
There is also a slowdown of growth for the case of constant matter supply due to the decrease of the concentration $\bar{c}(t)$ with time  (Eq.~\eqref{eq:conc1}).
In other words, a constant matter supply density $J$ does not provide enough material for droplets of increasing size such that their growth rate decreases.  
There is, however, no slowdown of growth in the case of constant supersaturation, where the droplet grows at a constant speed proportional to $(k \bar{c})$.

\section{Dynamics of two droplets}\label{sec-two}
\begin{figure*}[htp]
\centering
\includegraphics[width=1\linewidth]{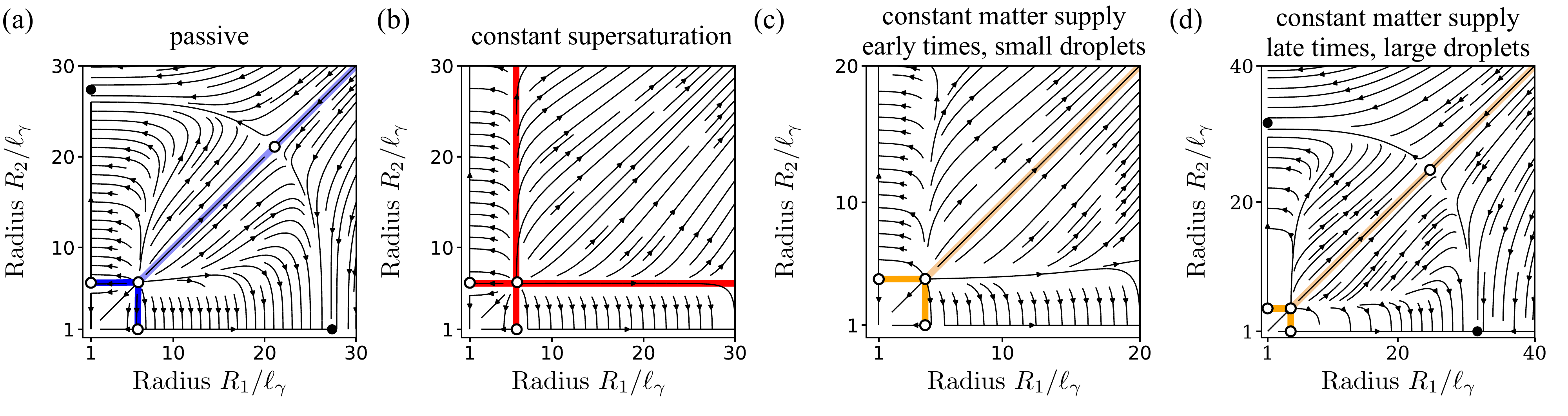}
    \caption{\textbf{Dynamics of two droplets in the diffusion-limited regime.} Phase portrait of two droplets with radii $(R_1, R_2)$ in the (a) passive, (b) constant supersaturation, and (c, d) the case with constant matter supply. 
    Black lines and arrows indicate the flow field lines $(\dot{R}_1,\dot{R}_2)$, and colored solid lines the separatrices. A separatrix splits the phase portraits into domains that differ in the asymptotic behaviour at long times. Solid disks represent stable fixed points, and open circles are unstable fixed points. (a) In the passive case, there are three domains of dynamics. Droplets with radii smaller than the critical radius  always shrink and dissolve, and there are two domains within which either $R_1$ or $R_2$ will not dissolve but reach the steady state. (b) For the constant supersaturation, there are three domains where one or both droplets shrink.
    The domain in the upper right corner describes the growth of both droplets. 
    This growth is indefinite, and there is no stable fixed point. 
    (c) At early times and for the constant supply, the dynamics are similar to those of the constant supersaturation. This similarity arises because droplet material is in excess, and the small droplets grow initially indefinitely. (d) At a late time and for the constant supply, the dynamics are similar to the passive case 
    since droplets grew to large sizes where the constant matter supply is growth-limiting.
    As a result, there are two stable fixed points and a single droplet persisting over large times. Results were obtained solving Eqs.~\eqref{eq:cons-law-discrete2-c},~\eqref{eq:cons-law-integralc} for the diffusion-limited regime with the parameters given in Table~\ref{tab-param2}.
    }
\label{fig3-diff-lim-1}
\end{figure*}
We next discuss the behaviour of two droplets ($N=2$) competing for the shared droplet material $\bar{c}$ in the diffusion- and interface-resistance-limited growth regime, in the presence of matter supply.
To this end, we solve the dynamical system of equations, in which each droplet $R_i$ evolves according to Eq.~\eqref{eq-drdt0} coupled via Eq.~\eqref{eq:cons-law-discrete}, and specified to the case $N=2$:
\begin{subequations}
\label{eq:cons-law-discrete2-all}
\begin{align}
\label{eq:dt_barc}
      \frac{\text{d}}{\text{d} t}  \bar{c}(t) &= -\frac{{4 \pi} c^\text{in,(0)}}{V_\text{sys}}
    \sum_{i=1,2}
    R_i(t)^2 \, \frac{\text{d}R_i(t)}{\text{d}t}  + J(t)
\,,\\
     \frac{\text{d}}{\text{d} t} R_i(t) &= \frac{k}{1 + \beta\left(R_i\left(t\right)\right)} \frac{1}{c^\text{in,(0)}}\Big(\bar{c}(t) -c^\text{eq}\left(R_i\left(t\right)\right)\Big)\,,
     \label{eq:cons-law-discrete2-c}
\end{align}
\end{subequations}
with the equilibrium interface concentration $c^\text{eq}(R_i)$ (Eq.~\eqref{eq:ceq_finite}) 
and the non-dimensional quantity $\beta$ (Eq.~\eqref{eq:beta}).
 
We solve Eqs.~\eqref{eq:cons-law-discrete2-all} for the two cases of supply and compare to the case without supply ($J = 0$) with the corresponding results shown in Fig.~\ref{fig3-diff-lim-1}(a-c). For the constant supply ($J =$ const.) and the passive ($J = 0$) case, we use the solution of the integrated Eq.~\eqref{eq:dt_barc} which reads
\begin{equation}
   \bar{c}(t) =  \bar{c}(0) - \frac{ 4\pi c^\text{in,(0)} }{3 V_\text{sys}}  \sum_{i=1,2} R_i(t)^3 + {J} \, t \,,
   \label{eq:cons-law-integralc}
\end{equation}
while for the constant supersaturation case, the coarse-grained background concentration is constant with $\bar{c} = \bar{c}(0)$.

The competition between two droplets can be represented as phase portraits, as used previously for chemically active emulsions~\cite{zwicker_suppression_2015}. 
Fig.~\ref{fig3-diff-lim-1} depicts the
dynamics of two droplet radii described by Eqs.~\eqref{eq:cons-law-discrete2-all} in the diffusion-limited growth. 
Stable and unstable steady states are shown as closed and open circles, respectively. 

For the passive case (see Fig.~\ref{fig3-diff-lim-1}(a)), there is a distinct unstable steady state $(R_\text{c},R_\text{c})$ with two large droplets of equal size, where $R_\text{c}$ is the critical radius.
When (at least) one of the droplets exceeds this critical radius,
the bigger one grows at the expense of the other one, leading to a single droplet in steady state. The corresponding stable steady state radius is set the total amount of droplet material in the system. 

The phase portraits of the cases with matter supply strongly differ to the passive case. Supplying matter using a  constant supersaturation, 
there is no stable fixed point for two droplets; see Fig.~\ref{fig3-diff-lim-1}(b).
Droplets larger than the critical radius will grow indefinitely. 

The case of constant matter supply (Fig.~\ref{fig3-diff-lim-1}(c)) is more subtle with a phase-portrait that is time-dependent. 
At early times, when droplets are smaller, the phase portrait is more similar to the case of constant supersaturation. 
Droplets can initially grow once both exceed the critical radius. The similarity relies on the smaller droplet growing seemingly unbounded when it takes up droplet material less than supplied by the constant matter supply.
At later times, when droplets are bigger, their growth gets limited by the matter supply rate. The results are stable fixed points similar to the passive case, where one of the droplets persists in the system. However, in contrast to the passive case, this single droplet is slowly growing, moving the fixed point to larger droplet sizes. 

We note that the phase portraits for interface-resistance-limited transport are similar. In particular, qualitative features such as the nullclines, fixed points, and the separatrix do not change. 

\section{Dynamics of the droplet size distribution  with matter supply}
\label{sec-emulsion}

This section discusses an emulsion with matter supply in the thermodynamic limit, where the total droplet number $N\to \infty$, and the system size $V_\text{sys} \to \infty$ with a finite droplet number density $n(t)=\lim_{N, V_\text{sys} \to \infty} N(t)/V_\text{sys}$. 
We derive the corresponding continuum theory for the droplet distribution function $\mathcal{N}(R,t)$ and the corresponding droplet size distribution function from the discrete model proposed in Sect.~\ref{eq:discrete_model}.
To this end, we define the 
droplet distribution function for an emulsion with $N$ droplets: 
\begin{equation}
\mathcal{N}(R,t) =  \lim_{N, V_\text{sys} \to \infty}
\frac{1}{V_\text{sys}}
\sum^{N(t)}_{i=1}  \delta\big(R-R_i(t)\big)\,.
\label{eqC2:left}
\end{equation}
In the continuum limit of the droplet distribution $\mathcal{N}(R,t)$, we consider the thermodynamic limit with droplet number density 
\begin{equation}
    n(t) = \int_0^\infty \text{d}R \, \, \mathcal{N}(R,t) \, ,
    \label{eqC2:number_cont}
\end{equation}
such that $\mathcal{N}(R,t) \text{d}R/n(t)$ is the normalized droplet size distribution function.

In the absence of nucleation and fusion of droplets, the droplet distribution obeys a continuity equation
\begin{equation}
\partial_t \mathcal{N}(R,t) + \partial_R\big(\dot{R}(R)\, \mathcal{N}(R,t) \big) = 0\,,
\label{eq:em-cont}
\end{equation}
and the droplet number density $n(t)$ can only change due to droplets vanishing through dissolution at $R=0$:
\begin{equation}
   \frac{\text{d}}{\text{d} t} n(t)  = -\left.\Big(\dot{R}(t) \mathcal{N}(R,t)\Big) \right|_{R=0}\,,
\label{eqC2:law_denistyn}    
\end{equation}
where we abbreviate $\dot R = {\text{d} R}/{\text{d} t}$. 
Furthermore, the $k$-th moment of the distribution is
\begin{equation}
   \langle R^k(t) \rangle = \frac{\int_0^\infty\text{d}R \, R^k(t)\,   \mathcal{N}(R,t) }{\int_0^\infty \text{d}R \,   \mathcal{N}(R,t) }\,,
\label{eqC2:kth_moment}
\end{equation} 
and we define the droplet phase volume fraction $\Phi(t)$: 
\begin{equation}
   \Phi(t) = 
   \int_0^\infty\text{d}R \, \frac{4\pi}{3} R^3(t) \, \mathcal{N}(R,t) \,.
\label{eq:C2:def_Phi}
\end{equation}
The gain and loss terms of the discrete model (Eq.~\eqref{eq:cons-law-discrete}), describing droplet growth and shrinkage, map in the continuum limit according to:
\begin{equation}
 \Phi(t) =\lim_{N,V_{\rm sys}\rightarrow \infty}
  \frac{1}{V_\text{sys}} \sum_{i=1}^{N(t)} \frac{4\pi}{3} R_i^3(t) %
  \,.
\end{equation}
Thus, the discrete dynamic Eqs.~\eqref{eq:cons-law-discrete} and \eqref{eq-drdt0} for the emulsion with matter supply take the form:
\begin{subequations}\label{eq:dynamic_barc_R}
\begin{align}
       \frac{\text{d}}{\text{d} t}  \bar{c}(t) &= -c^\text{in,(0)} \frac{\text{d}}{\text{d}t} \Phi(t) + J(t)\,,
    \label{eq:cons-law-cont}\\
     \frac{\text{d}}{\text{d} t} R(t) &= \frac{k}{1 + \beta(R)} \frac{1}{c^\text{in,(0)}}\Big(\bar{c}(t) - c^\text{eq}(R)\Big)\,,
    \label{eq-drdt00}
\end{align}
\label{eq:model-both}
\end{subequations}
with the evolution of the droplet size distribution given by Eq.~\eqref{eq:em-cont}.
In Eqs.~\eqref{eq:dynamic_barc_R},  we have used the definition of the droplet phase volume fraction (Eq.~\eqref{eq:C2:def_Phi}), the non-dimensional quantity $\beta(R) = k R/D^\text{out}$, and the Gibbs-Thomson relation for the equilibrium concentration $c^\text{eq}(R)$ at the droplet interface $R$.

The matter supply density $J(t)$ affects the shape of the droplet size distribution and its moments, such as the average radius or the standard deviation. In the following, we study emulsions with matter supply and compare the results to passive emulsions without matter supply ($J = 0$).

The LSW theory of Ostwald ripening governs the dynamics of passive emulsion~\cite{lifshitz_kinetics_1961,wagner_theorie_1961}, serving as an essential reference system for this work that discusses the effects of matter supply on emulsion dynamics.
Below, we summarize the key results of the LSW theory, which provides an analytical solution for the universal droplet size distribution reached at long times in a passive emulsion (no matter supply). 
The universal droplet size distribution can be obtained by separating the time-dependent function $g(t)$ and the distribution of the rescaled droplet size $h(\rho)\rho$ in the droplet distribution function:
\begin{equation}
    \mathcal{N}(R,t) = g(t)h(\rho)\rho\,, \quad \rho = R/R_\text{c}\,,
    \label{eq:separation-ansatz}
\end{equation} 
where $\rho$ is the droplet radius rescaled by the critical radius $R_\text{c}=\ell_\gamma/\varepsilon$ and $\mathcal{N}(R,t) R_\text{c}(t) \text{d}\rho/n(t)$ is the scaled droplet size distribution. 

For passive emulsions, the droplet size distribution broadens in time for both cases, 
diffusion-limited and interface-resistance-limited transport.  
In the case of diffusion-limited transport, the average radius follows
\begin{equation}
    \langle R(t) \rangle = \Bigg(\frac{4 D^\text{out} c^{(0)} \ell_\gamma}{9 c^\text{in,(0)}} t\Bigg)^\frac{1}{3}\,,
    \label{eqAvR:passive-diff}
\end{equation}
while for the interface-resistance-limited case,   
\begin{equation}
\langle R(t) \rangle = \frac{8}{9}\Bigg(\frac{ k c^{(0)} \ell_\gamma}{2 c^\text{in,(0)}} t\Bigg)^\frac{1}{2}\,.
\label{eqAvR:passive-int}
\end{equation}

The average radius $\langle R(t) \rangle$ rescaled by the critical radius $R_\text{c}(t)$ 
\begin{equation}
\label{eq:kappa_def_general}
    \kappa \coloneqq \frac{\langle R(t) \rangle}{R_\text{c}(t)} 
\end{equation}
turns out to be an important quantity that characterizes ripening dynamics of emulsions.
This quantity can be calculated from the distribution of the
rescaled droplet size $h(\rho)\rho$.
For passive emulsions governed by the LSW theory,
\begin{equation}
    \kappa = \frac{\int_0^\infty h(\rho) \rho^2 d\rho}{\int_0^\infty h(\rho) \rho\,  d\rho}\,.
\end{equation}
In the diffusion-limited regime, $\kappa = 1$, while in the interface-resistance-limited regime, $\kappa = 8/9$. The distribution of the rescaled droplet size $h(\rho)\rho$ in the passive case of the diffusion-limited transport reads~\cite{wagner_theorie_1961,lifshitz_kinetics_1961}: 
\begin{equation}
\begin{split}
h(\rho)\,\rho &= \rho^2\bigg(\frac{3}{3+\rho}\bigg)^\frac{7}{3} \Bigg(\frac{\frac{3}{2}}{\frac{3}{2}-\rho}\Bigg)^\frac{11}{3}
\exp{\Bigg(-\frac{\rho}{\frac{3}{2}-\rho}\Bigg)}\,,\\&\quad 0\leq\rho \leq \frac{3}{2}\,,
\end{split}
\label{eq:passive-distr-diff}
\end{equation}
and in the interface-resistance-limited regime, we have~\cite{wagner_theorie_1961}:
\begin{equation}
    h(\rho)\rho = \rho \bigg(\frac{2}{2-\rho}\bigg)^{5} \exp{\bigg(-\frac{3 \rho}{2-\rho}\bigg)}\,,\quad 0\leq \rho \leq2 \,.
    \label{eq:passive-distr-int}
\end{equation}
The time-dependent contribution to the droplet distribution $\mathcal{N}(R,t)$ (defined via Eq.~\eqref{eq:separation-ansatz}) is given by
\begin{equation}
    g(t) = \frac{3\Phi(0)}{4\pi R_\text{c}(t)^4 \int_0^\infty \rho^4 {h}(\rho) d\rho}\,,
\end{equation}
with the asymptotic solution of the critical radius in the diffusion-limited regime $R_\text{c}(t) = \big({4 D^\text{out} c^{(0)} \ell_\gamma} t \,/{9 c^\text{in,(0)}}\big)^{1/3}$, and in the interface-resistance-limited regime, $R_\text{c}(t) = \big({ k c^{(0)} \ell_\gamma} t\,/{2 c^\text{in,(0)}}\big)^{1/2}$.
In the following sections, we study the behavior of the droplet size distribution and its moments in the presence of matter supply, and compare them to the passive case corresponding to vanishing matter supply ($J = 0$).

\subsection{Constant supersaturation}

\begin{figure*}[tb]
    \centering
    \includegraphics[width=.9\linewidth]{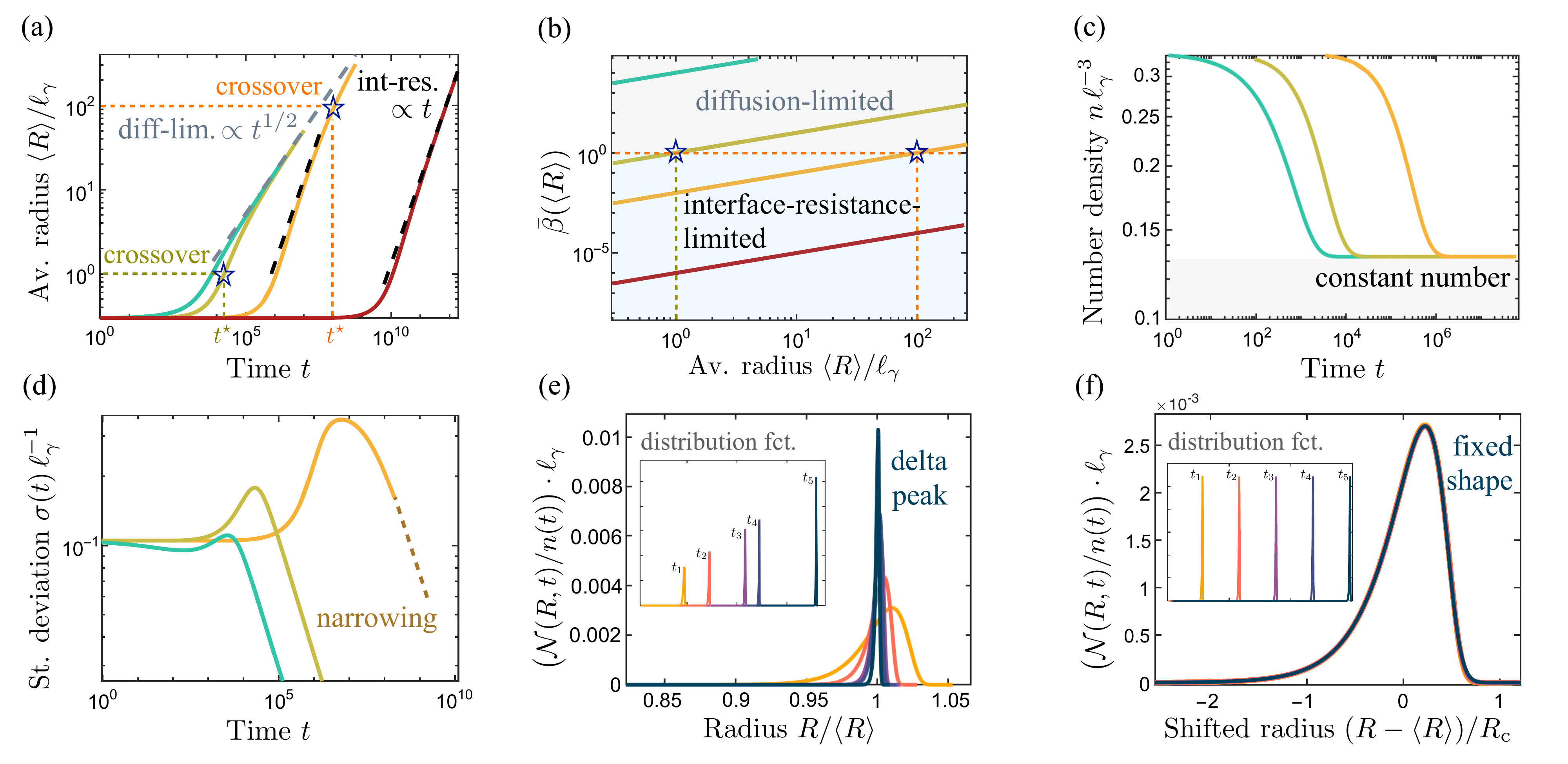}
    \caption{\textbf{Coarsening kinetics in an emulsion with constant supersaturation.} (a) Average radius as a function of time $t = t/\tau$, $\tau = \ell_\gamma^2/D^\text{out}$. The solid lines correspond to different initial choices of $\beta(\langle R(0)\rangle)$. Depending on the initial choice, the growth of emulsion is determined by diffusion- or interface-resistance-limited transport, with the average radius following $t^{1/2}$ or $t$ power law, respectively. The star marks the crossover between the two regimes for a particular initial condition. (b) The crossover from the interface-resistance to diffusion-limited regime happens when $\beta(\langle R \rangle) \geq 1$, indicated by the star. (c) After all droplets smaller than the critical radius have dissolved, the droplet number density saturates at a constant value. (d) After an initial regime of droplet dissolution, the standard deviation decreases. (e) In the diffusion-limited regime, the droplet size distribution function narrows towards a delta peak and is not universal. (f) In the interface-resistance-limited regime, the droplet size distribution function drifts with increasing velocity in space, maintaining a fixed shape and a constant standard deviation. Results were obtained solving Eqs.~\eqref{eq:em-cont}, \eqref{eqC3:dotR_cconst} for constant supersaturation $\varepsilon$, with the parameters given in Table~\ref{tab-param}.
    }
    \label{fig:const-eps}
\end{figure*}

Keeping the supersaturation, $\varepsilon={\bar{c}}/{c^{(0)}}-1$, constant requires that the matter supply density $J(t)$ balances the changes of the droplet phase volume fraction at all times such that the coarse-grained concentration value $\bar{c}$ does not change in time. This balance corresponds to the left-hand side of Eq.~\eqref{eq:cons-law-cont} being zero, i.e., $J(t)=c^\text{in,(0)}{\text{d}}\Phi(t)/{\text{d}t}$, with a constant $\bar{c}$ in Eq.~\eqref{eq-drdt00} and 
\begin{subequations}
\begin{align}
\frac{\text{d}}{\text{d} t} R(t) &= \frac{k}{1 + \beta(R)} \frac{c^{(0)}}{c^\text{in,(0)}}\bigg(\varepsilon - \frac{\ell_\gamma}{R}\bigg)\,.
\label{eqC3:dotR_cconst}    
\end{align}
\end{subequations}
Each droplet size $R(t)$ evolves independently in time (decoupled growth).
In other words, in an emulsion with constant supersaturation $\varepsilon$, the growth of every droplet is independent of the emulsion kinetics, and there is no competition for the droplet material among the droplets. Furthermore, a constant supersaturation $\varepsilon$ implies a constant critical radius $R_\text{c}=\ell_\gamma/\varepsilon$. Droplets with $ R > R_\text{c}$ at $t=0$ in $\mathcal{N}(R,t=0)$ will not dissolve and the droplet number density $n(t)=n(t=0)$ is conserved. 
Each droplet grows by absorbing droplet material from the fixed coarse-grained background concentration.

After droplets grew to sizes corresponding to the decreasing tail of ${\text{d}}R(t)/{\text{d} t}$ (see Eq.~\eqref{eqC3:dotR_cconst}), the droplet size distribution narrows. 
The mechanism of narrowing relies on 
larger droplets grow more slowly than smaller ones, which has been discussed for spatially heterogeneous quenches in emulsions~\cite{Weber_2017}.
The phenomenon of a narrowing droplet size distribution implies that the higher-order moments are small compared to the lower moments at each instance of time during narrowing. Therefore, we will focus on the average droplet radius and the radius standard deviation in the following.

Since droplets are decoupled from each other, and every droplet follows the growth law Eq.~\eqref{eqC3:dotR_cconst} independently, we can replace $R$ by the average radius $\langle R \rangle$.
The average radius thus follows 
\begin{equation}
     \frac{\text{d}}{\text{d} t} \langle R(t) \rangle \simeq \frac{k}{1 + \bar{\beta}(\langle R \rangle )} \frac{c^{(0)}\varepsilon}{c^\text{in,(0)}}\,,
\label{eqC3:dotR_cconst3}    
\end{equation}
where $\bar{\beta} = {k \langle R \rangle}/{D^\text{out}}$. The term $\ell_\gamma/\langle R \rangle$ can be neglected since at later times $\varepsilon \gg \ell_\gamma/\langle R \rangle$. Note that for constant supersaturation, droplets grow indefinitely with a positive growth rate ${\text{d}}\langle R(t) \rangle/{\text{d} t}>0$.

Eq.~\eqref{eqC3:dotR_cconst3} can be solved analytically. 
For diffusion-limited transport
($\bar{\beta} \gg 1$), compared to the passive solution Eq.~\eqref{eqAvR:passive-diff}, we find
\begin{equation}
    \langle R(t) \rangle = \Bigg(2\frac{D^\text{out} \, \varepsilon \, c^{(0)}}{c^\text{in,(0)}}\, t\Bigg)^\frac{1}{2}\,.
\label{eqC3:avR_cconst-diff}
\end{equation}
For the interface-resistance-limited transport, the average radius for constant supersaturation evolves according to
($\bar{\beta} \ll 1)$, 
\begin{equation}
 \langle R(t) \rangle = \frac{k \, \varepsilon \, c^{(0)}}{c^\text{in,(0)}}\, t\,.
    \label{eqC3:avR_cconst-int}  
\end{equation}
Please note the fast growth relative to the reference solution of the passive case (Eq.~\eqref{eqAvR:passive-int}).

We numerically solve Eqs.~\eqref{eq:em-cont} and \eqref{eqC3:dotR_cconst} for constant supersaturation $\varepsilon$. By varying the initial value of $\bar{\beta}(\langle R \rangle)$, for the same initial droplet radii, we distinguish between diffusion- and interface-resistance-limited regimes.
Fig.~\ref{fig:const-eps}(a) shows the results of the average radius for the different initial choices of $\bar{\beta}$ indicated in Fig.~\ref{fig:const-eps}(b). There is a crossover in the scaling of the average radius at time $t^\star$ from the linear scaling to a $t^{1/2}$ power law. Due to the increase of $\bar{\beta}$, the system is always governed by the diffusion-limited kinetics at late times (see Fig.~\ref{fig:const-eps}(b)).
Moreover, after the initial regime of dissolution of droplets that are smaller than the critical radius ($R < R_\text{c}$), droplets stop dissolving (Fig.~\ref{fig:const-eps}(c)).
Thus, the droplet number density $n(t)$ is indeed constant for longer times for the case where the supersaturation is kept constant.

The time change of the variance, $\sigma^2=\langle R^2 \rangle - \langle R \rangle^2$, can be expressed as
\begin{equation}
    \frac{\text{d}}{\text{d} t} \sigma^2 =  \frac{\text{d}}{\text{d} t} \langle R^2 \rangle - 2 \langle R \rangle  \frac{\text{d}}{\text{d} t} \langle R \rangle\,.
\label{eqC3:dt_var1}    
\end{equation}
Now we express the r.h.s. of the above equation using the continuity Eq.~\eqref{eq:em-cont}, the definition Eq.~\eqref{eqC2:kth_moment}, and the growth law Eq.~\eqref{eqC3:dotR_cconst}. Moreover, we  neglect the $(\ell_\gamma/R)$-term considering big enough droplets and assume constant number density $n$:
\begin{equation}
\begin{split}
    \frac{\text{d}}{\text{d} t} \sigma^2(t) &\simeq \frac{2}{n} \Bigg[ \Bigg(\int_0^\infty dR \, \frac{ k R}{1 + \beta(R)}\, \mathcal{N}(R,t)\Bigg) - \\
    &-\langle R \rangle \int_0^\infty dR \frac{k }{1 + \beta(R)}\, \mathcal{N}(R,t) \Bigg]\,,
\label{eqC3:sigma_both}
\end{split}
\end{equation}
where, after partial integration, we have assumed vanishing boundary terms at $R = 0,\infty$.

In the diffusion-limited regime, where for all droplets in the ensemble, $\beta(R)\gg 1$, we obtain:
\begin{equation}
\label{eq:dt_sigma}
    \frac{\text{d}}{\text{d} t} \sigma^2(t) \simeq 2 D^\text{out}\,\frac{c^{(0)}\varepsilon}{c^\text{in,(0)}} \Bigg(1-\langle R \rangle \bigg\langle \frac{1}{R} \bigg\rangle \Bigg)\,.
\end{equation}
Since $1/R$ for $R>0$ is a convex function, we use Jensen's inequality~\cite{j_l_w_v_jensen_sur_1906}.
For the inequality to be valid, the conditions satisfied by the droplet distribution $\mathcal{N}(R,t)$ are: 
$\mathcal{N}(R> 0,t) \geq 0$, otherwise $\mathcal{N}(R\leq 0,t) = 0$, the moments must be finite, and the droplet size distribution function normalizable. 
For the special case of $\mathcal{N}(R,t)$ being a delta function, the equality sign holds, while for all other distributions, the inequality is fulfilled.
It is only important that $1/R$ is convex, which is the case for $R>0$, implying $\langle R \rangle \langle 1/R\rangle \geq 1$. We conclude for the r.h.s. of Eq.~\eqref{eq:dt_sigma} that, due to Jensen's inequality, the variance (l.h.s.) decreases in time:
\begin{equation}
    \frac{\text{d}}{\text{d} t} \sigma^2(t) \leq 0\,.
\end{equation}
Consistent with the results depicted in Figs.~\ref{fig:const-eps}(d,e), the droplet size distribution function in the diffusion-limited regime for emulsions with constant supersaturation always narrows and approaches a delta function with time.

Now, let us turn to the case of interface-resistance-limited transport, where for all droplets in the ensemble, $\beta(R)\ll 1$.
In this case, Eq.~\eqref{eqC3:sigma_both} gives
\begin{equation}
    \frac{\text{d}}{\text{d} t} \sigma^2(t) = 0\,,
\end{equation}
meaning that the width of the distribution is steady. 
Indeed, the droplet size distribution in the interface-resistance-limited regime with constant supersaturation moves with a constant shape and drifts for increasing velocity in the $R$-space (see Fig.~\ref{fig:const-eps}(f)). Note that for cases initialized in the interface-resistance-limited regime with small enough radii ($\langle R\rangle \ll k/D^\text{out}$), the droplet distribution will eventually crossover to the diffusion-limited regime, transiting from a drifting distribution of a fixed shape to a narrowing droplet size distribution $\mathcal{N}(R,t)$.

\subsection{Constant matter supply}

For a constant matter supply, the coarse-grained background concentration $\bar{c}(t)$ and the droplet radii $R_i(t)$ are coupled in general (Eqs.~\eqref{eq:cons-law-cont}, \eqref{eq:C2:def_Phi}).
This coupling becomes quasi-static when matter is supplied at a rate much slower compared to droplet growth, such that droplets approximately absorb any excess, and there is hardly any supersaturation ($\varepsilon\simeq 0$, Eq.~\eqref{eq:const_supersaturation}). In this case, the time changes of the coarse-grained background concentration are quasi-static with $\text{d} \bar{c}/\text{d}t \simeq 0$,
leading to a dynamic equation for the droplet phase volume fraction $\Phi(t)$ defined in Eq.~\eqref{eq:C2:def_Phi}:
\begin{equation}
     c^\text{in,(0)}\frac{\text{d}}{\text{d}t} \Phi(t) \simeq J\,.
     \label{eq:cons-cont-const}
\end{equation}

\begin{figure*}[tb]
    \centering
    \includegraphics[width=0.9\linewidth]{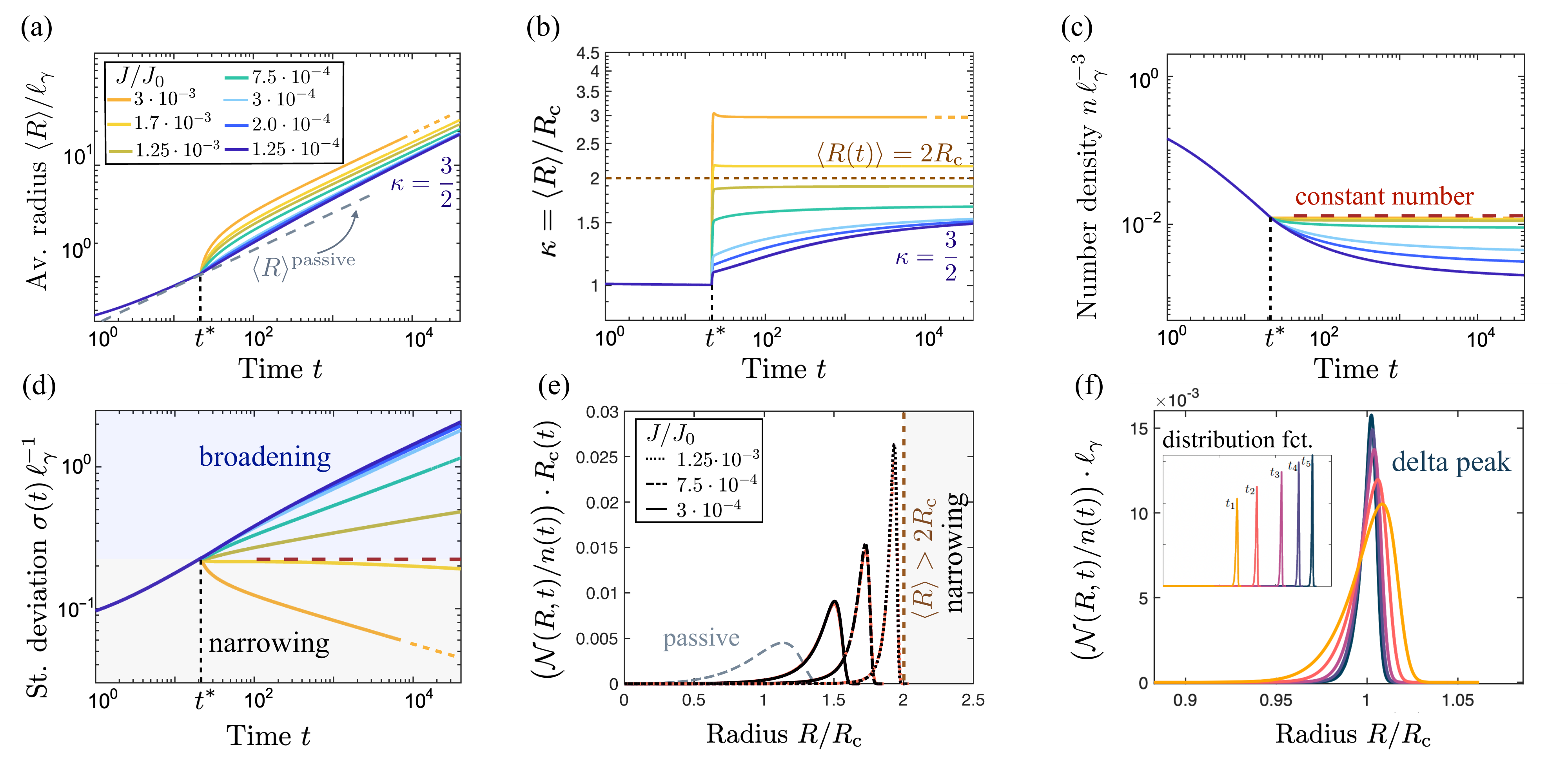}
    \caption{\textbf{Non-universal coarsening kinetics in an emulsion with constant matter supply in the diffusion-limited regime.} (a) Average radius for different values of constant supply density $J$. The supply is switched on at $t^*$, with time rescaled as $t \to t/\tau$, and $\tau = \ell_\gamma^2/D^\text{out}$. (b) The average radius rescaled by the critical radius for different values of supply. The droplet size distribution function narrows for $\kappa > 2$ (red dashed line). For the supply such that $\kappa < 3/2$, the coarsening converges to a broadening distribution function, for which $\kappa$ relaxes to $\kappa = 3/2$~\cite{vollmer_ripening_2014}. Since, in this limit, the asymptotic solution of the distribution function depends on the number density at the time the supply starts, we term it semi-universal. For $\kappa > 3/2$, its value depends on the matter supply $J/n(t^*) = 4\pi \ell_\gamma c^{(0)} D^\text{out}(\kappa -1)$. (c) Droplet number density converges to a constant value (red dashed line).  
    (d) Standard deviation crosses from the broadening to the narrowing regime with the increasing supply density $J$. The color code for the supply strength is indicated in (a).
    (e) The droplet size distribution function collapsed for a supply in the broadening regime via rescaling of the spatial coordinate by the critical radius. The choice of supply strength corresponds to the broadening regime. The color code represents different times for measuring the distribution function, while the line style corresponds to the strength of the supply density. 
    (f) The droplet size distribution function narrows towards a delta peak for a supply in the narrowing regime. The color code represents the measurement times, with yellow indicating early times and blue indicating late times. Results were obtained solving Eqs.~\eqref{eq:em-cont}, \eqref{eq:model-both} for different values of constant matter supply density $J$, with the parameters given in Table~\ref{tab-param}.
    }
    \label{fig:em-diff-lim}
\end{figure*}

\begin{figure*}[tb]
    \centering
    \includegraphics[width=0.9\linewidth]{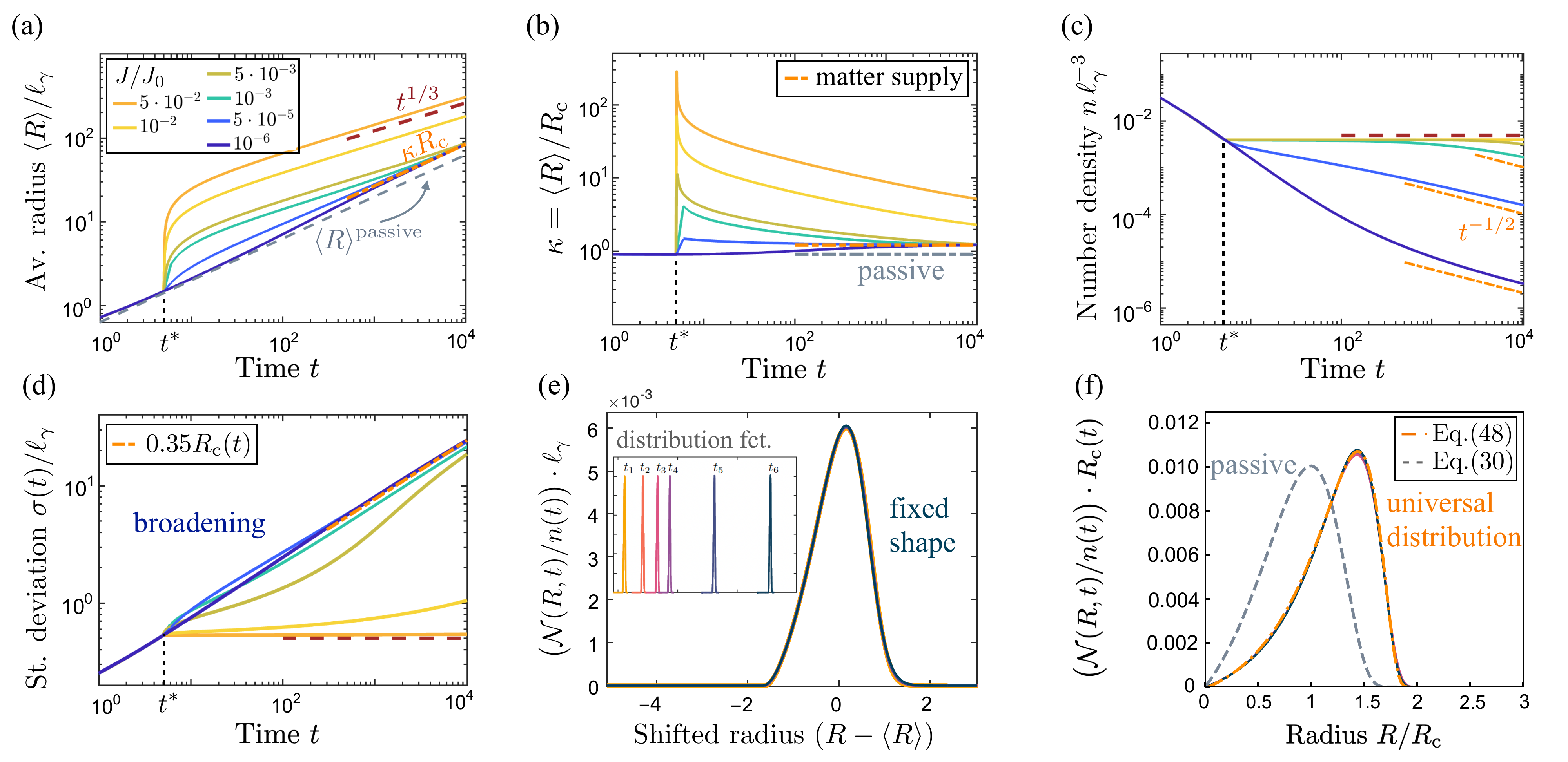}
    \caption{\textbf{Universal coarsening kinetics in an emulsion with constant matter supply in the interface-resistance-limited regime.} 
    (a) The average radius after an initial transient acceleration $\propto t^{1/3}$, converges to a new power law that is independent of the supply strength (orange dashed line, Eq.~\eqref{eqC3:avR_intflim}). The supply is switched on at $t^*$, with time rescaled as $t \to t/\tau$, and $\tau = \ell_\gamma^2/D^\text{out}$. 
    (b) The parameter $\kappa$ converges to a constant value that is supply independent, $\kappa \approx 1.19$. 
    It is above the passive value $\kappa = 8/9$ (gray-dashed). (c) The droplet number density follows the power law $\propto t^{-1/2}$ (orange dashed-dotted line). For a very high supply, the droplet number density is transiently constant (red dashed). 
    (d) The standard deviation is increasing for all supply strengths. The droplet size distribution function stays transiently constant in shape in the interface-resistance-limited regime and converges at late times to the universal behavior Eq.~\eqref{eq:sigma-int} with $R_\text{c}(t)$ in Eq.~\eqref{eq:dynRc-sol2} (orange-dashed).
    (e) Droplet size distribution function $\mathcal{N}(R,t) \ell_\gamma /n(t)$ at early times for a supply density such that droplet dissolution transiently ceases.
    Droplets in this regime are transiently decoupled, and the emulsion evolves with a constant standard deviation. 
    The distribution function drifts in space with a fixed shape.
    The color code represents the different times at which the distributions were calculated,  with yellow for early and blue for the late times. 
    (f) Late time asymptotic behavior of the droplet size distribution function, when the average radius converges to Eq.~\eqref{eqC3:avR_intflim}. In the regime of broadening, the scaled droplet size distribution is universal, $\mathcal{N}(R,t) R_\text{c}(t) \text{d}\rho/n(t) =  {h}(\rho)\rho\text{d}\rho/4$. For the passive case (gray dashed), we use Eq.~\eqref{eq:passive-distr-int}. The analytical solution matches the numerical time-series results. The solution for the passive case is indicated in the gray-dashed line. Results were obtained solving Eqs.~\eqref{eq:em-cont}, \eqref{eq:model-both} for different values of constant matter supply density $J$, with the parameters given in Table~\ref{tab-param}.}
    \label{fig:em-intlim}
\end{figure*}

In the following, we discuss the dependence of the coarsening kinetics on the matter supply density $J$. We will distinguish between the diffusion-limited regime ($\bar{\beta} \gg 1$), which was studied in the literature~\cite{vollmer_ripening_2014, clark_focusing_2011}, and the interface-resistance-limited transport ($\bar{\beta} \ll 1$). To distinguish the diffusion-limited from the interface-resistance-limited regimes,
we consider an ensemble of droplets with the non-dimensional quantity $\bar{\beta} = k\langle R \rangle/D^\text{out}$. 
We numerically solve Eqs.~\eqref{eq:em-cont} and \eqref{eq:model-both} for different values of constant matter supply density $J$ and using parameter values and initial droplet radii corresponding to the two regimes.

We first revisit the diffusion-limited regime~\cite{vollmer_ripening_2014, clark_focusing_2011}, where the droplet number density approaches a constant for long times (Fig.~\ref{fig:em-diff-lim}(c)). The average radius exhibits scaling behavior with time $\langle R \rangle \simeq A \, t^{1/3}$, where the prefactor $A$ depends on the matter supply density $J$ ~\cite{vollmer_ripening_2014}, see Fig.~\ref{fig:em-diff-lim}(a). For the standard deviation scaling law, both the prefactor and the exponent depend on the matter supply density $J$ ~\cite{vollmer_ripening_2014}, see Fig.~\ref{fig:em-diff-lim}(d). The droplet size distribution function changes from broadening to narrowing over time. The crossover occurs when ${J} \geq 4\pi n(t) D^\text{out} \ell_\gamma {c^{(0)}}$, corresponding to $\langle R(t) \rangle \geq 2 R_\text{c}(t)$, or $\kappa \geq 2$ (Fig.~\ref{fig:em-diff-lim}(b,d)) \cite{vollmer_ripening_2014, clark_focusing_2011}. 
As discussed in Ref.~\cite{vollmer_ripening_2014}, the average radius rescaled by the critical radius $\kappa$ (Eq.~\eqref{eq:kappa_def_general}) relaxes to $\kappa = 3/2$ at late times but only for the values of supply such that $\kappa(t^*) \leq 3/2$ (Fig.~\ref{fig:em-diff-lim}(e)).
For the cases with matter supply such that $\kappa > 3/2$, we confirm that the emulsion evolves towards a monodispersed distribution peaked at $\langle R \rangle$, which depends on the matter supply (blue vs. green lines in Fig.~\ref{fig:em-diff-lim}(b), and Ref.~\cite{vollmer_ripening_2014}). In the regime of narrowing droplet size distribution ($\kappa > 2$), the droplet size distribution is not universal and depends on the initial conditions (Fig.~\ref{fig:em-diff-lim}(f) and Ref.~\cite{vollmer_ripening_2014}). 

We now discuss the interface-resistance-limited regime, in which we find a universal droplet size distribution which broadens with time (Fig.~\ref{fig:em-intlim}(d,e)). The universal distribution function can be obtained analytically using a method similar to that used in the LSW theory for the passive case~\cite{lifshitz_kinetics_1961,wagner_theorie_1961}.
Substituting in Eq.~\eqref{eqC2:law_denistyn} the growth law for a single droplet (Eq.~\eqref{eq-drdt00}) in the interface-resistance-limited regime (${\beta} \ll 1$), 
the time evolution of the droplet number density can be written as: 
\begin{equation}
    \frac{\text{d}}{\text{d} t} n(t)= - \left. \Bigg( k \frac{c^{(0)} \ell_\gamma}{c^\text{in,(0)}} \frac{1}{R} \,\mathcal{N}(R,t) \Bigg)\right|_{R=0}\,.
    \label{eq:n-int-cont}
\end{equation}
Analogously to Eq.~\eqref{eq:separation-ansatz}, we again use an ansatz that the droplet distribution can be expressed as $\mathcal{N}(R,t) = g(t) {h}(\rho)\rho$, as a product of a time-dependent function $g$, and a function ${h}(\rho)\rho$ of the rescaled radius $\rho = R/R_\text{c}$. To ensure that the number density $n(t)$ (Eq.~\eqref{eq:n-int-cont}) is finite for small $R$, we impose ${h}(\rho)\rho \simeq \rho$ for small $\rho$, with ${h}(\rho = 0) = 1$.

We now determine the time-dependent part $g(t)$ of the droplet distribution. 
Using the rate of change of the droplet phase volume fraction (Eq.~\eqref{eq:cons-cont-const}), the definition of the droplet phase volume fraction $\Phi(t)$ (Eq.~\eqref{eq:C2:def_Phi}) and the separation ansatz $\mathcal{N}(R,t) = g(t) {h}(\rho) \rho$, we obtain:
\begin{equation}
    g(t) = \frac{3 \big( \Phi(0) + t \,J/c^\text{in,(0)}\big)}{4\pi R_\text{c}(t)^4 \int_0^\infty \rho^4 {h}(\rho) d\rho}\,.
\label{eq:em:defg}
\end{equation} 
Substituting in Eq.~\eqref{eq:n-int-cont} the definition of the droplet number density $n(t)$ (Eq.~\eqref{eqC2:number_cont}), together with the separation ansatz $\mathcal{N}(R,t) = g(t) {h}(\rho) \rho$ and the solution for $g(t)$ (Eq.~\eqref{eq:em:defg}), we find the dynamics of the critical radius in the presence of constant matter supply:
\begin{equation}
    \frac{\text{d}}{\text{d} t} R_\text{c} = \frac{k c^{(0)} \ell_\gamma}{3 c^\text{in,(0)} \alpha} R_\text{c}^{-1} + \frac{J/c^\text{in,(0)}}{3\big(\Phi(0) + J t/c^\text{in,(0)} \big)} R_\text{c}\,,
    \label{eq:dynRc}
\end{equation}
where $\alpha = \int_0^\infty {h}(\rho) \rho d\rho$ is the normalization of the distribution of the rescaled droplet size ${h}(\rho)\rho$. The solution of Eq.~\eqref{eq:dynRc} for the critical radius 
\begin{equation}
\begin{split}
R_\text{c}(t)^2 &=
\bigg(R_{\text{c}}(0)^2 -\frac{6 K \Phi(0)}{J/c^{\text{in,(0)}}}\bigg)\Bigg(1+\frac{{J t}/{c^{\text{in,(0)}}}}{\Phi(0)}\Bigg)^\frac{2}{3}
\\ & \quad +\frac{6 K \Phi(0)}{J/{c^{\text{in,(0)}}}} + 6 K t \,,
\end{split}
\label{eq:dynRc-sol}
\end{equation}
where $R_\text{c}(0)$ is the critical radius at $t=0$. Moreover, we abbreviated $K={k c^{(0)} \ell_\gamma}/{(3 c^\text{in,(0)} \alpha)}$. We are interested in the asymptotic behavior of the critical radius at long times $t\rightarrow\infty$. The solution $R_\text{c}(t)^2$ has three contributions of the order of: $t^{2/3}, t^0, t$. 

In the limit $t\rightarrow\infty$, we can determine the asymptotic solution 
\begin{equation}
R_\text{c}(t) = \sqrt{6Kt}\,\Big(1+\mathcal{O}\big(t^{-{1}/{3}}\big)\Big)\,.
\label{eq:dynRc-sol2}
\end{equation}
Using the abbreviation $K$, Eq.~\eqref{eq:dynRc-sol2} gives the asymptotic critical radius $R_\text{c}(t) = (2 k c^{(0)} \ell_\gamma t/ (c^\text{in,(0)} \alpha))^{1/2}$ (see Appendix~\ref{app:critical_rad} how to obtain the asymptotic solution). Note that Eqs.~\eqref{eq:C2:def_Phi},\eqref{eq:cons-cont-const},\eqref{eq:em:defg} and the solution of $R_\text{c}(t)$ give a constraint for the distribution function to be normalizable. Due to a linear increase of the droplet phase volume fraction, and $t^{1/2}$ scaling of the critical radius, it follows that $\int_0^\infty d\rho\, \rho^4 h(\rho) = \text{const.}$ Furthermore, Eqs.~\eqref{eq:em:defg},\eqref{eqC2:number_cont} and the solution of $R_\text{c}(t)$ imply that the droplet number density is decreasing.

Let us now calculate the droplet size distribution in the asymptotic limit.   
We solve the continuity equation~\eqref{eq:em-cont} for $\Phi(0)\,c^\text{in,(0)}\ll  Jt$,  
using Eq.~\eqref{eq:em:defg}, and the asymptotic solution of the critical radius. This leads to a differential equation for the distribution of the normalized droplet size (see Appendix~\ref{app:critical_rad2} for detailed calculation):
\begin{equation}
\label{eq:h_rho_eq}
{h}^\prime(\rho) \big(\rho^2 \alpha^{-1} -\rho + 1 \big)+{h}(\rho) \big(3 \rho \alpha^{-1} - 1\big) = 0\,.
\end{equation}
A well-defined solution to Eq.~\eqref{eq:h_rho_eq} exists on a finite support $\rho\in[ 0, 2]$, for $\alpha = 4$ (for detailed calculation see Appendix~\ref{app:solution-dist-func}), and reads:
\begin{equation}
    {h}(\rho)\rho = \rho\bigg(\frac{2}{2-\rho}\bigg)^3\,\exp\bigg(-\frac{\rho}{2-\rho}\bigg)\,,\quad 0 \leq \rho \leq 2\,.
    \label{eq:int-lim-sol-prev}
\end{equation}

For the value of $\alpha = 4$, the critical radius $R_\text{c}(t) = (k c^{(0)} \ell_\gamma t/ (2 c^\text{in,(0)}))^{1/2} 
$ and is identical to the passive case. This result is consistent with the reported convergence of the critical radius to the corresponding passive case (see Appendix~\ref{app:critical_rad}, Fig.~\ref{fig:C2_Rc_const}(b)). We note that the distribution of the rescaled droplet size $h(\rho) \rho$ (Eq.~\eqref{eq:int-lim-sol-prev}) is independent of physical parameters (matter supply, transport coefficients, etc.) and time, satisfying the properties of a universal distribution.

In summary, for interface-resistance-limited transport and constant matter supply, we found a closed-form expression of the droplet distribution $\mathcal{N}(R,t)$
\begin{equation}
    \mathcal{N}(R,t) =\frac{\tilde{J} \,t \,  }{R_\text{c}(t)^4 }\, \frac{\rho}{(2-\rho)^3}  \,\exp\bigg(-{\frac{\rho}{2-\rho}}\bigg), \quad \rho=\frac{R}{R_\text{c}}\,,
    \label{eq:int-lim-sol}
\end{equation}
where the scaled matter supply rate $\tilde{J} = 6J /(\pi c^\text{in,(0)}\int_0^2 \rho^4 h(\rho) d\rho )$.
Using Eq.~\eqref{eq:int-lim-sol-prev}, we have $\tilde{J} = -J\big(\pi c^\text{in,(0)}\left( 32 + 56 e \,\mathrm{Ei}(-1) \right)\big)^{-1}$, where $\mathrm{Ei}(-1)=- \int_{1}^\infty dt \,e^{-t}/t\approx 0.219$ is the exponential integral. The scaled droplet size distribution $\mathcal{N}(R,t)R_\text{c}(t)\text{d}\rho/n(t) = h(\rho)\rho\text{d}\rho/4$, in the asymptotic regime, is universal and independent of matter supply density $J$.

Let us return to how this universal distribution emerges in time. Fig.~\ref{fig:em-intlim}(a-f) shows how the droplet size distribution and its moments change in time for different matter supply densities $J$. We compare the analytical and numerical solutions to the passive case without matter supply. 
When switching on the matter supply at time $t^*$, the average radius $\langle R \rangle$ strongly increases, transiting to an intermediate power-law scaling with $\langle R \rangle \propto t^{1/3}$ (Fig.~\ref{fig:em-intlim}(a), red dashed).
This initial increase is more pronounced for larger $J$. 
On long times, the average radius approaches a value equal for all $J$ but different from the passive system. 
The rescaled average radius, $\kappa=\langle R \rangle /R_\text{c}$, approaches a supply-independent value that can be calculated analytically.  Using the definition of the average radius (Eq.~\eqref{eqC2:kth_moment}) and the separation ansatz:
\begin{equation}
   \kappa = \frac{\int_0^2 \rho^2 {h}(\rho)d\rho}{\int_0^2 \rho {h}(\rho)d\rho} \,.
\label{eqC3:avR_intflim}    
\end{equation}
Note that $\kappa=\langle \rho\rangle _{h(\rho)\rho}$ can be understood as an average of the rescaled radius $\rho$ using the distribution of the rescaled droplet size, $h(\rho)\rho$.
Substituting the solution $h(\rho)$ (Eq.~\eqref{eq:int-lim-sol-prev}) above, we find $\kappa = -2 e \text{Ei}(-1) \approx 1.19$, which is indicated as orange dashed line in Fig.~\ref{fig:em-intlim}(b). The average radius is thus $\langle R(t)\rangle \simeq  1.19 \big(k c^{(0)} \ell_\gamma/2c^\text{in,(0)} t\big)^{1/2}$ (orange dashed line in Fig.~\ref{fig:em-intlim}(a)).

In the intermediate $t^{1/3}$-scaling regime of the average radius,
we find that the number density of droplets $n(t)$ is conserved and the standard deviation constant (Fig.~\ref{fig:em-intlim}(c,d)).
We can understand these results in the quasi-static limit (Eq.~\eqref{eq:cons-cont-const}), where the droplet phase volume fraction scales as $\Phi(t) \propto J\, t $. 
The general definition of the average volume $\langle V \rangle = \Phi(t)/n(t)$ can be expressed in terms of $\Phi(t)$ and $n(t)$.
Since the droplet size distribution function has a constant standard deviation in the transient regime, we can assume that $\langle R \rangle \propto \langle V \rangle^{1/3}$. Thus, we see that when $\langle R \rangle \propto t^{1/3}$, the droplet number density $n(t)=\text{const.}$ and thereby is conserved within the intermediate $t^{1/3}$-scaling regime. 

The arrest of dissolution can also be understood from the transient dynamics of the coarse-grained background concentration $\bar c$. In Appendix~\ref{app:quasi-static-conc}, specifically Eq.~\eqref{eq:barc-Jc-app}, we give the solution of the coarse-grained background concentration in the quasi-static limit, $\text{d} \bar{c}/\text{d}t \simeq 0$, in the interface-resistance-limited regime. 
It determines that in this regime, the critical radius is given by
\begin{equation}
R_\text{c}(t) = \frac{\langle R(t)^2\rangle}{\langle R(t)\rangle + J(8\pi\,n(t) k c^{(0)} \ell_\gamma)^{-1}}
\label{eq:Rc-Jc}
\end{equation}
and the growth law
\begin{equation}
\frac{\text{d}}{\text{d} t} R(t) = {k} \frac{ c^\text{(0)}\ell_\gamma}{c^\text{in,(0)}}\Bigg(\frac{\langle R(t)\rangle}{\langle R(t)^2\rangle}
+\frac{J \big( 8\pi \,n(t) k c^{(0)} \ell_\gamma\big)^{-1}}{\langle R(t)^2\rangle}- \frac{1}{R}\Bigg)\,,
\label{eq:dR-full}
\end{equation}
is governed by the dynamics of the droplet size distribution and its moments, as well as the matter supply density. Depending on which term dominates the growth, droplets can be coupled or grow independently. 

When the supply is introduced at time $t^*$, the sudden decrease of the critical radius (Eq.~\eqref{eq:Rc-Jc}) due to the matter supply density $J$, transiently diminishes the dissolution of droplets.
Thus, droplets are decoupled from each other, and the growth law for $\text{d}R/\text{d}t$ (Eq.~\eqref{eq:dR-full}) is transiently dominated by the $J$ term rather than the distribution of the droplet sizes, which becomes transiently constant (Fig.~\ref{fig:em-intlim}(d)).
This behavior is, however, restricted to short times (Fig.~\ref{fig:em-intlim}(c)), and as the droplet phase volume fraction $\Phi(t)$ increases further, the number density is not constant anymore, and the average radius crosses to the new and universal $t^{1/2}$-power-law behavior.
The coarsening is then dominated by the first term on the r.h.s. of Eq.~\eqref{eq:dR-full}. 
At all times, droplets are coupled to each other and grow depending on all other droplets. 
The broadening size distribution marks this regime, while a narrowing and monodispersed emulsion, as in the diffusion-limited regime, corresponds to the decoupling of droplets.

In the long-time $t^{1/2}$-scaling regime of the average radius, the droplet number density decays with the power-law $n(t)\propto t^{-1/2}$ (Fig.~\ref{fig:em-intlim}(c)).
This finding can be confirmed analytically by substituting Eq.~\eqref{eq:int-lim-sol} and Eq.~\eqref{eq:dynRc-sol2} into the definition of the droplet number density (Eq.~\eqref{eqC2:number_cont}):
\begin{equation}
     n(t) = t^{-\frac{1}{2}}\,\frac{\tilde{J}}{2}\Bigg( \frac{kc^{(0)}\ell_\gamma}{2c^\text{in,(0)}}\Bigg)^{-\frac{3}{2}} \,.
\end{equation}
Note that for the passive case in the interface-resistance-limited regime, 
the droplet number density follows $n(t)\propto t^{-3/2}$~\cite{wagner_theorie_1961}.

The analytical solution of the droplet distribution $\mathcal{N}(R,t)$ can also be used to determine higher moments such as the standard deviation $\sigma = \sqrt{\langle R^2\rangle - \langle R \rangle^2}$ of the droplet size distribution on long time-scales, which is given by a critical radius times a constant:
\begin{equation}
\sigma(t) = R_\text{c} (t)\Bigg(\frac{\int_0^2 \rho^3 {h}(\rho) d\rho}{\int_0^2 \rho {h}(\rho)d\rho} - \kappa^2 \Bigg)^\frac{1}{2}\,.
\end{equation}
The constant of multiplication after the evaluation of the integrals is
\begin{align}
2 \sqrt{-e \text{Ei}(-1) (e \text{Ei}(-1)+4)-2}\approx 0.35 \,,
\label{eq:sigma-int}
\end{align}
such that $\sigma(t) \approx 0.35\, R_\text{c}$.
This analytical result is independent of $J$, and agrees with the results of the standard deviation for different strengths of the matter supply density (orange dashed line in Fig.~\ref{fig:em-intlim}(d)).
The droplet size distribution is broadening, and on large time scales, the standard deviation, as well as the average radius, follows a supply-independent and universal power-law behavior, which is different from that of a passive emulsion. 
We note that the dependence on the critical radius, $\sigma(t) \propto R_\text{c}(t)$, is the same as in passive emulsions. 
However, it differs in terms of the prefactor, which results from the altered distribution of the rescaled droplet size (Eq.~\eqref{eq:int-lim-sol-prev}) compared to passive systems.

\begin{figure*}[tb]
    \centering
    \includegraphics[width=\linewidth]{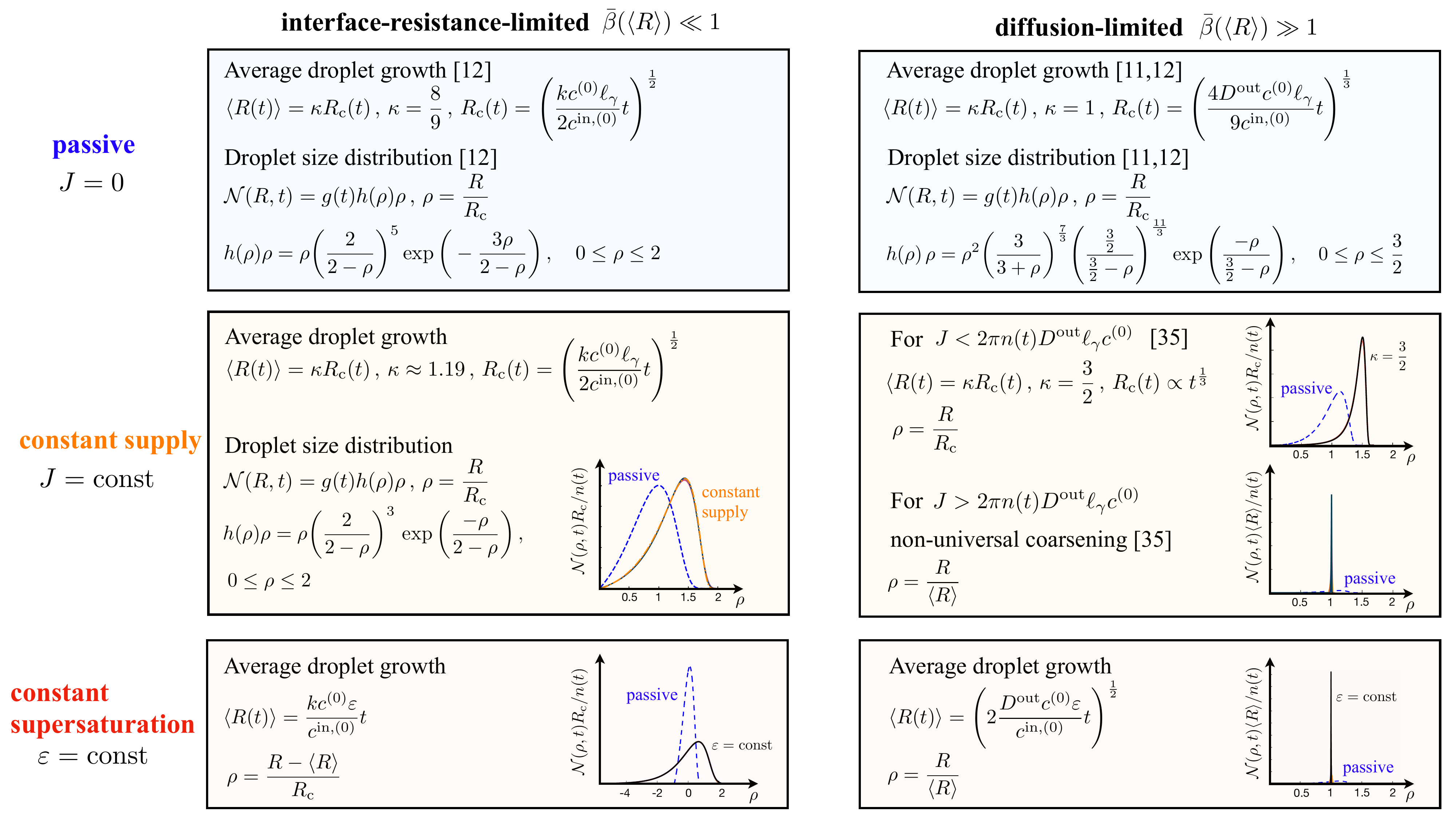}
    \caption{\textbf{Summary of the universality classes in emulsions with matter supply}. The parameter $\bar\beta(\langle R \rangle)$ distinguishes the diffusion- and interface-resistance-limited transports. For the three cases of matter supply, we show the types of universality classes that emerge in the system and the asymptotic solution governing the behaviour of the emulsion. For passive emulsions without matter supply ($J = 0$), the emulsion coarsens according to the LSW theory. It is essential to note that for interface-resistance-limited processes, in the presence of diffusion, a transition to the diffusion-limited regime occurs at very late times. This is because as the average radius increases, so does $\bar\beta(\langle R \rangle)$. The universal classes are defined for asymptotic times when the system is in its respective growth regime and relaxed to its asymptotic solution. 
    We found universal coarsening kinetics for constant matter supply, and $\bar{\beta} \ll 1$. The emulsion evolves at late times according to the universal solution Eq.~\eqref{eq:int-lim-sol-prev}, independent of matter supply. The distribution function is broadening, and the critical radius converges to the passive solution. 
    In the diffusion-limited regime,  only for $\kappa < 3/2$, the coarsening kinetics is semi-universal since it depends on the number density at the time of the supply switch $t^*$, such that if $J < 2\pi n(t^*) \ell_\gamma c^\text{out} D^\text{out}$, $\kappa$ relaxes to $\kappa = 3/2$. For a supply such that $\kappa > 3/2$, the coarsening is non-universal and depends on $J$. For $J = 4\pi n(t^*)\ell_\gamma c^\text{out} D^\text{out}$, there is a crossover from a broadening to a narrowing distribution function~\cite{vollmer_ripening_2014, clark_focusing_2011}. 
    This crossover indicates a loss of coupling between the droplets and a convergence to a monodispersed emulsion, with a delta peak centered at the average radius.
    For constant supersaturation, we found novel coarsening kinetics, where droplets evolve according to a new power law behavior $\langle R\rangle \propto t^{1/2}, t$, for $\bar{\beta}(\langle R\rangle) \ll 1, \bar{\beta}(\langle R\rangle) \gg 1$, respectively. 
    In the interface-resistance-limited regime, the distribution drifts in space with a fixed shape and increasing velocity. 
    In the diffusion-limited regime, it is centered as a delta peak around the average radius $\langle R \rangle$. 
    The case of constant supersaturation is the limit of independent growth of all droplets, which is in contrast to coupled growth in the passive case.}    \label{fig:final}
\end{figure*}

In summary, the key finding of this section is that emulsions with matter supply evolve according to a universal coarsening kinetics 
in the regime of interface-resistance-limited growth
(Fig.~\ref{fig:em-intlim}(b,f)). 
An intermediate coarsening regime exists when the number density is constant for intermittent times due to a higher supply density $J$, for which the critical radius decreases to a value at which droplets stop dissolving. 
In this regime, the average radius evolves with a $t^{1/3}$ power law, and the distribution function drifts along the axis of the droplet radius $R$ with a fixed shape and an increasing velocity (Fig.~\ref{fig:em-intlim}(e)). 
However, at late times, the coarsening becomes independent of the initial conditions and universal for all supply strengths. 
At late times, droplets will grow with the same universal power law and converge to the same universal distribution function (Fig.~\ref{fig:em-intlim}(a,f)). 
This strongly contrasts the non-universal and supply-dependent coarsening kinetics for the diffusion-limited transport in emulsions with constant matter supply (see Fig.~\ref{fig:em-diff-lim} and Ref.\cite{vollmer_ripening_2014}).

\section{Discussion}

In this work, we developed the theory for the kinetics of emulsions with matter supply. 
It extends the seminal LSW theory for passive systems for diffusion- and interface-resistance-limited transport by matter supply. 
Moreover, we generalize
more recent work with constant matter supply~\cite{clark_focusing_2011, vollmer_ripening_2014} to interface-resistance-limited transport and cases where a chemostat maintains a constant supersaturation. 
In a nutshell, we discuss how the matter supply affects the coarsening kinetics of emulsions.

The main universal regimes with and without matter supply, and for both diffusion-limited
and interface-resistance-limited transport
are summarized in Fig.~\ref{fig:final}. The value of the non-dimensional quantity $\bar\beta(\langle R \rangle) = {k \langle R \rangle}/{D^\text{out}}$ distinguishes diffusion-limited transport ($\bar\beta \gg 1$) and interface-resistance-limited transport ($\bar\beta \ll 1$) for an ensemble of droplets of average radius $\langle R \rangle$.
Here, the key kinetic parameters are the interfacial conductivity $k$    
and the diffusion coefficient $D^\text{out}$ of the background field between the droplets. 
The matter supply $J$ determines the dynamics of the emulsions, particularly for time-dependent fluxes that maintain a constant supersaturation. For passive systems, the matter supply $J$ is zero.  

The dynamics with and without matter supply differ largely by the coupling between the droplets via the dilute phase. 
Without matter supply ($J=0$),  passive emulsions are completely coupled through the background concentration of the dilute phase. Matter supply $J$ can transiently or fully decouple droplets depending on the rate-limiting transport process, either 
limited by diffusion in the dilute phase 
or resisting transport through the interface.  
In the interface-resistance-limited regime with a constant matter supply, droplets are weakly coupled, leading to scaling laws on intermediate time scales and transient decoupling, marked by a constant standard deviation, Fig.~\ref{fig:em-intlim}(d,e). On long time scales, when droplets become large enough, they affect the background concentration of the dilute phase and thus become coupled again, marked by a broadening distribution, Fig.~\ref{fig:em-intlim}(d,f). For the diffusion-limited regime, droplets either remain weakly coupled and the distribution continues to broaden or, for sufficiently high supply, droplets grow independently of one another, and the distribution continuously narrows, Fig.~\ref{fig:em-diff-lim}(d-f). Thus, for constant matter supply, over long times, the slowest growth process controls the behavior of the emulsion.

However, when the supersaturation is maintained constant, the decoupling of droplets persists over time, with a constant droplet number density for both diffusion-limited and interface-resistance-limited transport.

There is universal coarsening behavior in the presence of matter supply that depends on the non-dimensional quantity $\bar\beta$ and the matter supply density $J(t)$ (Fig.~\ref{fig:final}). 
We find novel scaling laws for the moments of the droplet size distribution and a universal (time- and parameter-independent) 
shape of the droplet size distribution function. 
In the interface-resistance-limited regime, we describe a novel universal coarsening kinetics for constant matter supply. We derived the distribution function, which is independent of matter supply and universal, and different from passive emulsions.

The critical radius converges to the one for the passive system (Fig.~\ref{fig:C2_Rc_const}(b)).   
The emulsion behavior is different in the diffusion-limited regime (see Fig.~\ref{fig:em-diff-lim} and Refs.~\cite{clark_focusing_2011, vollmer_ripening_2014}), in which the droplet kinetics depend on the initial conditions and the strength of the matter supply.

Finally, for constant supersaturation, we found non-universal droplet size distributions independently of the rate-limiting process. 
For interface-resistance-limited transport, the droplet size distribution drifts with a fixed shape in space, and in the diffusion-limited case, the droplet size distribution narrows towards a delta peak. 

Our theory on emulsion dynamics with matter supply is relevant for chemically fuelled emulsions. 
The matter supply $J$ can effectively be realized by a chemical reaction in which a precursor molecular component undergoes a fuel-driven chemical reaction. 
Previous studies have reported accelerated Ostwald ripening in chemically fuelled emulsions, with the average radius increasing with the same power-law behavior as the passive system in diffusion-limited transport~\cite{tena2021accelerated}. However, the prefactor of the power law was shown to be determined by the chemical reaction rates. 
In light of our work, this corresponds to the prefactor set by the matter supply $J$, with $\langle R \rangle \simeq A \,  t^{1/3}$, where the prefactor $A$ depends on $J$~\cite{vollmer_ripening_2014}, see Fig.~\ref{fig:em-diff-lim}(a). 
At early times, this supply rate was approximately constant when the precursors were in excess, since the chemical reaction was shown to be well-modeled by a first-order chemical reaction~\cite{schwarz_parasitic_2021,tena-solsona_non-equilibrium_2017}.
In short, accelerated ripening results from a matter supply of droplet material. 

We hypothesize that our theory on emulsion dynamics with matter supply could apply to the growth of biomolecular condensates in living cells. 
Active biochemical regulation is known to change kinetic timescales of emulsions in cells~\cite{sundararajan2025multiscalegrowthkineticsmodel}. 
Recent experimental evidence shows that the nucleolus average volume grows proportional to $t^3$ in \textit{Caenorhabditis elegans}~\cite{KodanWeber2025}, corresponding to a linear scaling of the nucleolus average radius. 
This scaling is consistent with emulsion dynamics at fixed supersaturation. 
Ref.~\cite{KodanWeber2025} proposed a model that supported the idea that condensate growth is driven by rRNA transcription, which is a form of matter supply. 
More generally, due to ongoing cell protein production, biomolecular condensates can be exposed to supersaturated concentration levels that are approximately constant ($\varepsilon = \text{const}$). 
This setting enables droplet growth that is significantly faster than that of passive systems without a matter supply. 
This accelerated growth law could enable biological cells to rapidly grow large numbers of condensates by bringing nearly all nucleated condensates to a mature size.

\begin{acknowledgements}
We are grateful for the insightful discussions with T.\ Tushar-Dutta, E.\ Ilker.
We are also grateful for the discussions with J.\ Boekhoven,  M.\ Tena-Solsona, and P.\ S.\ Schwarz on experiments with actively fuelled emulsions. 
C.\ Weber acknowledges the 
European Research Council (ERC) for financial support under the European Union's Horizon 2020 
research and innovation programme (``Fuelled Life'' with Grant agreement No.\ 949021).
\end{acknowledgements}

\appendix

\section{Derivation of the coarse-grained background concentration dynamics}
\label{app:calc-barc}

The concentration outside the droplets follows a diffusion equation,
\begin{equation}
\partial_t c^\text{out}(\boldsymbol{r},t) = - \nabla \cdot \boldsymbol{j}^\text{out}\,,
\label{eq1:main}
\end{equation}
where $\textbf{j}^\text{out} = -D^\text{out} \nabla c^\text{out}(\textbf{r},t)$, and  $\partial_t c^\text{in} = 0$, such that $\boldsymbol{j}^\text{in} = 0$. The boundary condition for the flux is 
\begin{equation}
\left. \boldsymbol{j}^\text{out} \cdot \boldsymbol{n} \right|_{\partial V} = \frac{J(t)}{S} \, ,
\end{equation}
where $J(t)$ is the in general time-dependent
matter supply rate.

Moreover, 
$S$ is the surface area of the box that corresponds to the boundary of the emulsion. We assume that in the far field, the concentration outside is constant in space, such that we can introduce the coarse-grained homogeneous concentration outside:
\begin{equation}
    \bar{c}(t) = \frac{1}{V_\text{sys}-V_\text{d}(t)} \int_{V_\text{sys}-V_\text{d}(t)} d^3 r \, c^\text{out}(\boldsymbol{r},t)\,.
\end{equation}
Using the boundary condition
\begin{equation}
    \frac{\text{d}}{\text{d} t} \bar{c}(t) = \frac{J(t)}{V_\text{sys}-V_\text{d}(t)}- \frac{4\pi  c^\text{in,(0)}}{V_\text{sys}-V_\text{d}(t)} \sum_i^{N(t)} R_i^2 \frac{\text{d}R_i}{\text{d} t}\,,
\end{equation}
for $V_\text{sys} \gg V_\text{d}(t)$, we obtain
\begin{equation}
    \frac{\text{d}}{\text{d} t} \bar{c}(t) =  \frac{J(t)}{V_\text{sys}}- \frac{4\pi  c^\text{in,(0)}}{V_\text{sys}} \sum_i^{N(t)} R_i^2 \frac{\text{d}R_i}{\text{d} t}\,,
\end{equation}
where $ {J(t)}/{V_\text{sys}}$ has units of concentration per time.

\section{Derivation of the interface-resistance-limited growth law}
\label{app-intlim-dRdt}

Here, we derive the relationship between the interface fluxes and the chemical potential differences across the interface. 
To this end, we consider a reference frame in which the interface is at rest.
Irreversible thermodynamics suggests a non-linear relationship between chemical potentials and the flux at the interface $R$~\cite{hubatsch_transport_2024}:
\begin{equation}
    \left. j\right|_R = \bar{k} \Big(e^{\frac{\mu^\text{in}}{k_\text{B}T}} - e^{\frac{\mu^\text{out}}{k_\text{B}T}}\Big)\,,
\label{eg-j-with_k}  
\end{equation} 
where the origin of the non-linear (exponential) dependence on the chemical potentials $\mu^\text{in/out}$ can be argued similarly to chemical reactions~\cite{van_kampen_nonlinear_1973}. 
The coefficient $\bar{k}$ has units $\left[\bar{k}\right] = (\text{concentration}) \cdot (\text{length/time})$ and characterizes the relaxation of the chemical potentials at the interface. 
We can write the chemical potential $\mu^\text{in/out}$ in both phases in terms of the activity coefficient $\gamma^\text{in/out}$ of both phases at equilibrium using 
\begin{equation}
\label{eq:mu_def}
 \mu^\text{in/out} = \mu^{\text{in/out},0} + k_\text{B} T \log{(\gamma^\text{in/out} c^\text{in/out})} \, .    
\end{equation}
At phase equilibrium, the partition coefficient 
\begin{equation}  \label{eq:P_def}
P\coloneqq\frac{c^\text{in,eq}(R)}{c^\text{out, eq}(R)}
\end{equation} 
can thus be expressed as
\begin{equation}
    P =  \frac{\gamma^\text{out}}{\gamma^\text{in}}\,.
\end{equation} 

In the following, 
we consider the case where the concentration inside is
approximately equal to the equilibrium concentration inside  ($c^\text{in} \simeq c^\text{in,eq}(R)$).
Substituting the chemical potentials (Eq.~\eqref{eq:mu_def}) into the flux through the interface (Eq.~\eqref{eg-j-with_k}), we obtain
\begin{equation}
\begin{split}
    \left.j\right|_R &= \bar{k} \gamma^\text{in}\bigg( c^\text{in,eq}(R) - \frac{\gamma^\text{out}}{\gamma^\text{in}} c^\text{out}(R)\bigg) \\
    &= \bar{k} \gamma^\text{in} P \bigg( c^\text{out,eq} (R) - c^\text{out}(R)\bigg) \, ,
\end{split}
\end{equation}
where used the definition of the partition coefficient (Eq.~\eqref{eq:P_def}) to obtain the last line.
We introduce the speed of relaxation at the interface $R$,
\begin{equation}
  k \coloneqq \bar{k}\,  \gamma^\text{in} P  \, , 
\end{equation}
and express the normal flux of droplet material evaluated outside and at the interface in the lab frame as
\begin{equation}    \left.j^\text{out}\right|_R  = k\Big(c^\text{eq}(R) - c(R)\Big)\,,
\end{equation}
where $c^\text{eq}(R)$ is the outside equilibrium concentration, which is given by the Gibbs-Thomson relation (Eq.~\eqref{eq:ceq_finite}). For brevity, we have dropped the superscript `out' for the concentrations and introduced it for the flux to indicate the equilibrium inside conditions.

\section{Single and two droplets in the interface-resistance-limited regime} 
\label{app-intlim2}
\begin{figure}[htb!]
    \centering
    \includegraphics[width=0.95\linewidth]{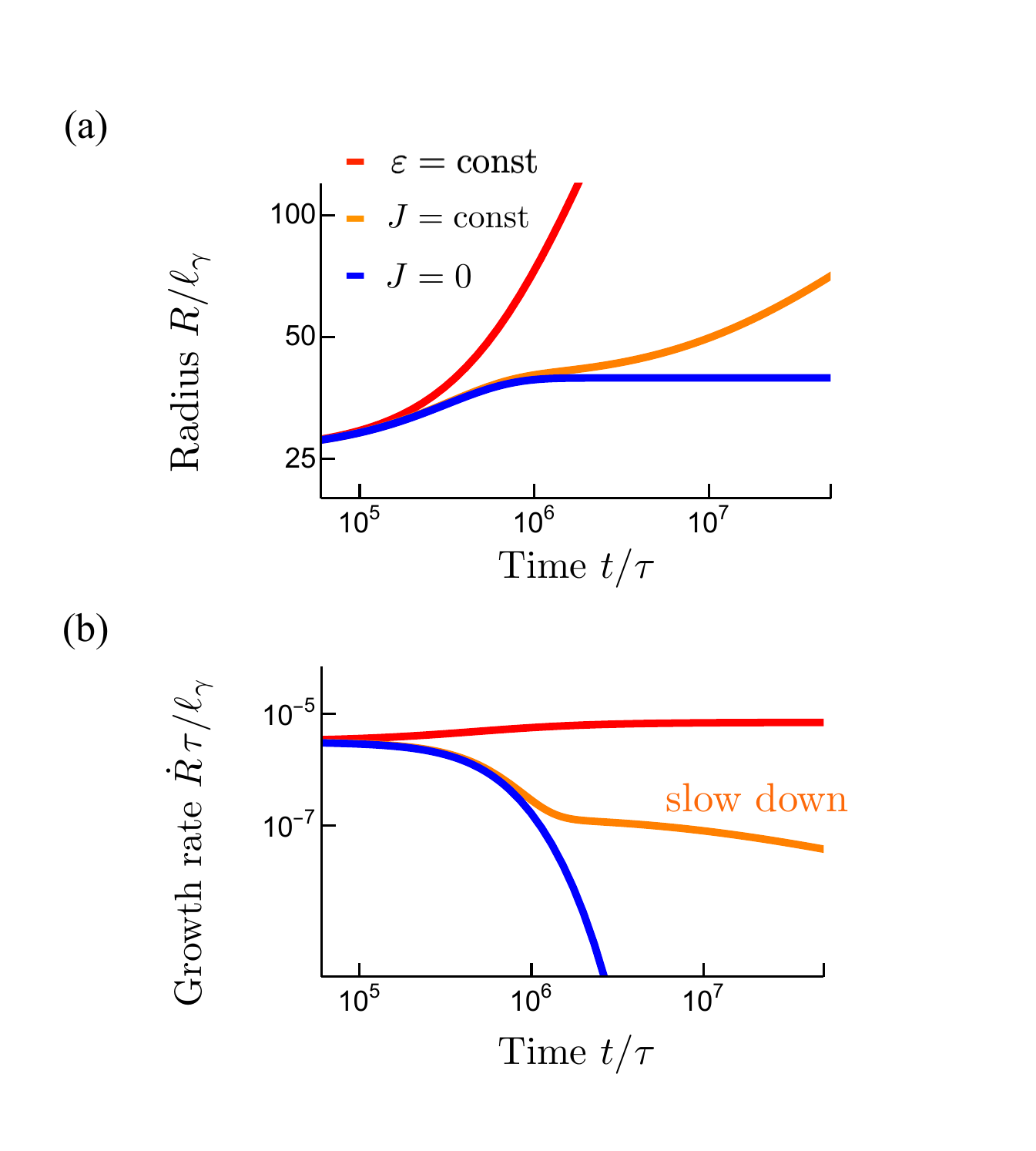}
    \caption{\textbf{Single droplet in a system coupled to a material reservoir. Results for the interface-resistance-limited regime.} (a) Growth of the droplet for the three supply cases. In the presence of matter supply, the radius grows indefinitely, while in the passive case, the growth ceases close to equilibrium. $\tau = \ell_\gamma/k$ in the log-log plot, $\ell_\gamma$ denotes the capillary length. (b) Growth speed of the droplet radius $R$ as a function of time in the log-log plot. For the constant supersaturation, the growth speed plateaus at a constant value. Results were obtained solving Eqs.~\eqref{eq:growth1} for the interface-resistance-limited regime with the parameters given in Table~\ref{tab-param1}.} 
    \label{int-lim1}
\end{figure}

Fig.~\ref{int-lim1} depicts analogous results to Fig.~\ref{fig2-diff-lim} for the interface-resistance-limited regime. 

\section{Asymptotic scaling of the critical radius}
\label{app:critical_rad}
\begin{figure*}[htb!]
    \centering
    \includegraphics[width=0.95\linewidth]{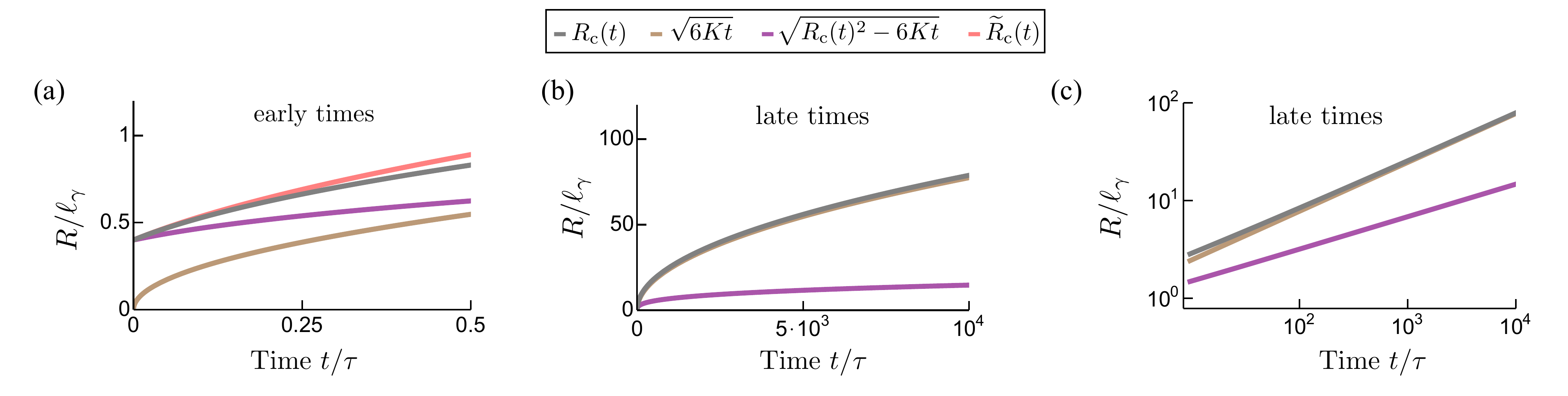}
    \caption{\textbf{Early and late time behavior of the critical radius.} The solution $R_\text{c}(t)$, Eq.~\eqref{eq:dynRc-sol}, is shown in gray, the leading order term $\sqrt{6Kt}$, which is also the $t\rightarrow\infty$ approximation, Eq.~\eqref{eq:dynRc-sol2}, in brown, the $t^{1/3}$ contribution, Eq.~\eqref{eq:dynRc-term3}, in purple, and the $t\simeq 0$ approximation $\widetilde{R}_\text{c}(t)$, Eq.~\eqref{eq:dynRc-early}, in pink.
    The chosen parameter values are consistent with the numerical simulation of emulsion kinetics, Fig.~\ref{fig:em-intlim}, with $\Phi(0) = 0.1$, $R_\text{c}(0)/\ell_\gamma = 0.4$, $J \tau/c_\text{in}^{(0)} =10^{-6}$ and $K \tau/\ell_\gamma^2 =10^{-4}$, where $\tau = \ell_\gamma/k$. (a) At early times, $\widetilde{R}_\text{c}(t)$ successfully approximates the solution Eq.~\eqref{eq:dynRc-sol}. (b) At late times, the solution converges to the leading order term approximation $\sqrt{6Kt}$ (c) Log-log plot to visualize the long time asymptotic behaviour.} 
    \label{int-lim1-Rc}
\end{figure*}

In the main text, we considered the long-time asymptotic behavior of the critical radius $R_\text{c}(t)$ determined by  Eq.~\eqref{eq:dynRc-sol}. 
In this section, we discuss the behavior of this equation, particularly its scaling behavior at early and late times.

Fig.~\ref{int-lim1-Rc} shows the full solution of Eq.~\eqref{eq:dynRc-sol} (gray) and compares its evolution to the leading order term $\sqrt{6Kt}$ (brown), Eq.~\eqref{eq:dynRc-sol2}, and the square-root of the remaining term (purple): 
\begin{equation}
\begin{split}
&R_\text{c}(t)^2-6Kt = \frac{6 K \Phi(0)}{J/{c^{\text{in,(0)}}}}\\
&\quad +\Bigg(1+\frac{{J t}/{c^{\text{in,(0)}}}}{\Phi(0)}\Bigg)^{2/3}\bigg(R_{\text{c}}(0)^2 -\frac{6 K \Phi(0)}{J/c^{\text{in,(0)}}}\bigg) \, . 
\end{split}   
\label{eq:dynRc-term3}    
\end{equation}
We note that the $t^{1/3}$ contribution is only relevant at early times (Fig.~\ref{int-lim1-Rc}(a)), and at long times Eq.~\eqref{eq:dynRc-sol} converges to $\sqrt{6Kt}$, Fig.~\ref{int-lim1-Rc}(b,c). 

We now discuss the behavior of the critical radius at early times ($t\simeq 0$). Using the binomial expansion gives
\begin{equation}
     \Bigg(1+\frac{{J t}/{c^{\text{in,(0)}}}}{\Phi(0)}\Bigg)^{2/3}
       =  1+\frac{2}{3}\frac{{J t}/{c^{\text{in,(0)}}}}{\Phi(0)}+\mathcal{O}(t^2)\,.
\end{equation} 
Inserting the expansion in Eq.~\eqref{eq:dynRc-sol}, the critical radius at early times can be written as 
\begin{equation}
\widetilde{R}_\text{c}(t) = \sqrt{ R_\text{c}(0)^2 +2t\bigg(K+R_\text{c}(0)^2\frac{J/c^\text{in,(0)}}{3\Phi(0)}\bigg) +\mathcal{O}(t^2)}\,,
\label{eq:dynRc-early}
\end{equation}
where $K={k c^{(0)} \ell_\gamma}/{(3 c^\text{in,(0)} \alpha)}$. We note that $\widetilde{R}_\text{c}(t)$ is an approximation of the critical radius ${R}_\text{c}(t)$ at early times. 

While at early times the $t^{1/2}$ order also dominates the dynamics of the critical radius, the effect of matter supply is non-negligible. This approximation works very well at early times (pink line in Fig.~\ref{int-lim1-Rc}(a)). 

\begin{figure}[htb]
    \centering
    \includegraphics[width=0.8\linewidth]{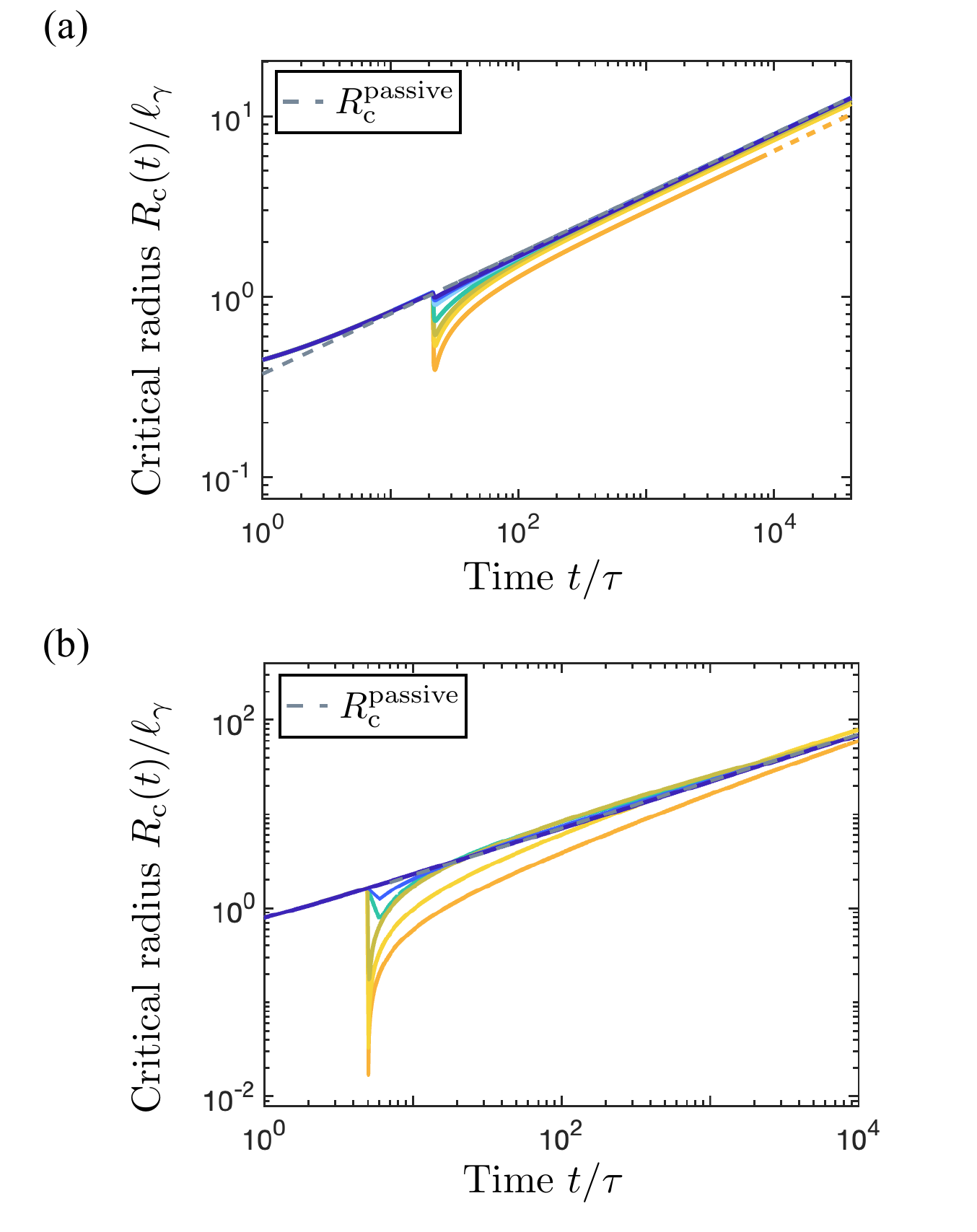}
    \caption{\textbf{Convergence of the critical radius in emulsions with constant matter supply.} The critical radius for emulsions with matter supply converges to the passive solution (gray dashed line). It is consistent with the results in Eq.~\eqref{eq:dynRc-sol2}. (a) Diffusion-limited regime. The orange and yellow solid lines correspond to the emulsion that is narrowing, and thus, which is not expected to converge to the passive solution. (b) Interface-resistance-limited regime. The orange solid line corresponds to the emulsion with a transiently constant standard deviation, for which the critical radius had not yet converged to the passive solution.
    Results were obtained solving Eqs.~\eqref{eq:em-cont} and \eqref{eq:model-both} for different values of constant matter supply density $J$, with the parameters given in Table~\ref{tab-param}. The values of $J$ and the color code in (a) correspond to Fig.~\ref{fig:em-diff-lim} and in (b) to Fig.~\ref{fig:em-intlim}.}
    \label{fig:C2_Rc_const}
\end{figure}

As discussed in the main text, at late times the critical radius converges to $R_\text{c} = \sqrt{6Kt}$ (Eq.~\eqref{eq:dynRc-sol2}), which is independent of matter supply density $J$; see brown line in Fig.~\ref{int-lim1-Rc}(b).
Furthermore, numerical results for the interface-resistance-limited regime show that the critical radius converges to the critical radius of the passive emulsion (which is valid for $\alpha = 4$ in the definition of $K$); see Fig.~\ref{fig:C2_Rc_const}(b). 

\section{Derivation of the differential equation for the rescaled distribution of the droplet sizes}
\label{app:critical_rad2}
In the following, we want to show the derivation of Eq.~\eqref{eq:h_rho_eq}. Using the separation ansatz $\mathcal{N}(R,t) = g(t) {h}(\rho) \rho$, with $\rho = R/R_\text{c}$, the first term of the continuity equation~\eqref{eq:em-cont}, $\partial_t \mathcal{N}(R,t)$ is:
\begin{equation}
    \partial_t \mathcal{N}(R,t) = {h}(\rho)\rho \frac{\text{d}g(t)}{\text{d}t}+ g(t) {h}^\prime(\rho)\rho \frac{\text{d}\rho}{\text{d}t} + g(t) {h}(\rho)\frac{\text{d}\rho}{\text{d}t}\,,
\end{equation}
where ${h}^\prime(\rho) = \text{d}{h}(\rho)/\text{d}\rho$.
Using the definition of $g(t)$ Eq.~\eqref{eq:em:defg} for the choice $\Phi(0)c^\text{in,(0)} \ll t J$, and $\text{d}\rho/\text{d}t = -\rho R_\text{c}^{-1} \text{d}R_\text{c}/\text{d}t$, it follows:
\begin{equation}
\begin{split}
    \partial_t \mathcal{N}(R,t) &= {h}(\rho) \rho\bigg(-4g(t)R_\text{c}^{-1}(t)\frac{\text{d}R_\text{c}(t)}{\text{d}t}+\frac{g(t)}{t}\bigg)\\
    &- g(t)\rho R_\text{c}^{-1}(t)\frac{\text{d}R_\text{c}(t)}{\text{d}t}\bigg({h}^\prime(\rho)\rho+{h}(\rho)\bigg)\,.    
\end{split}
\label{app:diffh:dtN1}
\end{equation}
We use the dynamics of the critical radius, $R_\text{c}^{-1}\text{d}R_\text{c}/{\text{d}t} = KR_\text{c}^{-2}+{t}^{-1}/3$ , where $K = kc^{(0)}\ell_\gamma/(3c^\text{in,(0)} \alpha)$ and $\Phi(0)c^\text{in,(0)} \ll t J$.
In this limit, we obtain the asymptotic solution, $R_\text{c}(t)^2=6Kt$, which no longer depends on $J$. Eq.~\eqref{app:diffh:dtN1} then gives:

\begin{equation}
     \partial_t \mathcal{N}(R,t) = -3K R_\text{c}(t)^{-2} g(t)\bigg(3{h}(\rho) \rho+{h}^\prime(\rho)\rho^2\bigg)\,.
\label{app:diffh:dtN2}     
\end{equation}
Using the growth law Eq.~\eqref{eq-drdt00} in the interface-resistance-limited regime, with $R_\text{c}=\ell_\gamma c^{(0)}/(\bar{c}(t)-c^{(0)})$, the second term of the continuity equation~\eqref{eq:em-cont}, $\partial_R(\text{d}R/\text{d}t\, \mathcal{N}(R,t))$ is:
\begin{equation}
\begin{split}
    &\partial_R\bigg(\frac{\text{d}R(t)}{\text{d}t} \mathcal{N}(R,t)\bigg) \\
     & \quad =3 \alpha K R_\text{c}^{-2} g(t)\Bigg({h}(\rho) +\rho {h}^\prime(\rho) - {h}^\prime(\rho)\Bigg) \,.
\end{split}
\label{app:diffh:dRN}     
\end{equation}
Taking the sum of Eqs.~\eqref{app:diffh:dtN2}, \eqref{app:diffh:dRN}, and dividing both sides by $-3KR_\text{c}^{-2}\alpha^{-1}g(t)$ gives Eq.~\eqref{eq:h_rho_eq}. We note that the separation of the rescaled radius $\rho$ and time $t$ was successful since there is no explicit time dependence in the equation above.

\section{Solution of the differential equation for the distribution of the rescaled droplet size}
\label{app:solution-dist-func}

To solve Eq.~\eqref{eq:h_rho_eq} for $h(\rho)$, we need to find the range of $\alpha$ for which the physical solution exists.
We first rewrite Eq.~\eqref{eq:h_rho_eq} as:
\begin{equation}
    \frac{\text{d}h(\rho)}{h(\rho)} = - \text{d}\rho \frac{3\rho\,\alpha^{-1}-1}{\rho^{ 2}\alpha^{-1}-\rho+1}\,.
\end{equation}
We want to find the physical solution of $h(\rho)$, which describes the distribution of droplet sizes during coarsening. For the problem of coarsening, we can assume that the physical solution of $h(\rho)$ exists on a finite support, with the left bound set by $\rho = 0$ and the upper bound by the maximum size of the droplets. 

On the finite support, there should be no singularity. We thus integrate both sides of the equation above on a finite support:
\begin{equation}
\begin{split}
    \ln\bigg(\frac{h(\rho)}{h(0)}\bigg) &= -\frac{3}{2}\int_0^{\rho} \text{d}\rho^\prime \frac{2\rho^\prime\,\alpha^{-1}-1}{\rho^{^\prime 2}\alpha^{-1}-\rho^\prime+1}\\
    &\quad -\frac{1}{2}\int_0^{\rho} \text{d}\rho^\prime \frac{1}{\rho^{^\prime 2}\alpha^{-1}-\rho^\prime+1}\,,
\end{split}    
\end{equation}
where an upper bound of the supporter has still to be determined. 
 
We have performed the decomposition of the fraction in the integrand. 
The first integral can be easily integrated, and it gives:
\begin{equation}
    h(\rho)-h(0) = |Q(\rho)|^{-\frac{3}{2} }\exp{\bigg(-\frac{1}{2}\int_0^{\rho} \text{d}\rho^\prime \frac{1}{Q(\rho^\prime)}\bigg)}\,,
\label{eq:app:Qdiffh1}
\end{equation}
for $h(\rho) > 0$.
The discriminant of the quadratic term $Q(\rho)={\rho^{2}\alpha^{-1}-\rho+1}$, is $\Delta = \sqrt{1-4\alpha^{-1}}$, and two roots $\rho_{2,1} = (1\pm\Delta)/2\alpha^{-1}$, where $\rho_{2}>\rho_{1}$.
For the finite support to be well-defined, $h(\rho)$ must vanish at $\rho=\rho_{2,1}$. We thus need to analyse the solution near the roots.

Here, we discuss the solutions of Eq.~\eqref{eq:app:Qdiffh1} for the following three cases of $\alpha$ and $\Delta$: (i) $0 < \alpha < 4$, $\Delta < 0$, (ii) $\alpha > 4$, $\Delta > 0$ (iii) $\alpha = 4$, $\Delta = 0$. These are exactly the regimes arising from the quadratic structure of the denominator, and leading to a finite support of the integral. We will show:
Case (i) contradicts the assumption of a finite support, with the support being infinite, and the droplet distribution not normalizable.
Case (ii) leads to a solution with higher-order derivatives diverging close to the upper bound, which is inconsistent with the solution having no singularity within the support. The only physical solution consistent with the assumption of a finite support and the condition of decreasing droplet number density is case (iii). 

(i) For $0 < \alpha < 4$, $\Delta < 0$, and there are no real roots of $Q$. Both terms on the r.h.s. of Eq.~\eqref{eq:app:Qdiffh1} never diverge, and no algebraic poles emerge from these factors. This contradicts the assumption of finite support, as the solution never vanishes for any upper bound, and the support is infinite. The distribution function is thus not normalizable.

(ii) For $\alpha > 4$, $\Delta > 0$, the first term leads to a multiplicative algebraic factor, with a power-law divergence. For the term in the exponent, we use partial fraction decomposition, which, after integration, leads to:

\begin{equation}
 \frac{h(\rho)}{\text{C}} = \alpha^{\frac{3}{2}}\, |(\rho - \rho_2)|^{-\frac{3}{2}-\frac{1}{2\Delta}} \cdot|(\rho - \rho_1)|^{-\frac{3}{2}+\frac{1}{2\Delta}}\,,
\end{equation}
where $\text{C}$ contains all integration constants. Since $0<\rho_1<\rho_2$, the finite range is determined by $\rho\in[ 0 ,\rho_1]$.

Let us analyse the local behavior near the root $\rho\rightarrow\rho_1$, such that $\delta_1\equiv   |\rho-\rho_1| \rightarrow 0$:
\begin{equation}
 \frac{h(\rho)}{\text{C}} \rightarrow \alpha^{\frac{3}{2}}\, |\rho_1 - \rho_2|^{-\frac{3}{2}-\frac{1}{2\Delta}}\cdot {\delta_1}^{-\frac{3}{2}+\frac{1}{2\Delta}}\,,\quad \rho \rightarrow\rho_1\,.
\end{equation}
For $\alpha > 4.5$, there is a power-law divergence at $\rho = \rho_1$, and the solution blows up (analogous calculation gives a divergence at the larger root $\rho_2$ for every $\alpha >4$). For the higher order derivatives, it follows that near the root $\rho\rightarrow\rho_1$:
\begin{equation}
 \frac{h^{(n)}(\rho)}{\text{C}} \rightarrow \alpha^{\frac{3}{2}}\, |\rho_1 - \rho_2|^{-\frac{3}{2}-\frac{1}{2\Delta}}\cdot {\delta_1}^{-\frac{3}{2}-n+\frac{1}{2\Delta}}\,,\quad \rho \rightarrow\rho_1\,.
\end{equation}
The range of $\alpha>4$, for which the higher order derivatives are well-defined close to the root, decreases with the order $n$. 

There are different lines of argument for finding the physical value of the separation constant $\alpha$. Lifshitz and Slyozov~\cite{lifshitz_kinetics_1961} argued that

the only stable solution is when $h(\rho)$ and all its derivatives go to zero at $\rho_1$, so that to the right of $\rho_1$ the solution is identically zero~\cite{lifshitz_kinetics_1961}. This corresponds to case (iii) when the two roots are identical, $\rho_{1}=\rho_2$. Another line of argument is presented in Ref.~\cite{bray_theory_1994}, which uses the fixed point analysis of $\text{d}\rho/\text{d}t$, $\rho =R/R_\text{c}$. In our case, ${\text{d}R}/{\text{d}t} = kc^{(0)}\ell_\gamma(R_\text{c}^{-1}-R^{-1})/c^\text{in,(0)}$ and $R_\text{c}(t) = (2kc^{(0)}\ell_\gamma t/\alpha)^{1/2}$ lead to:
\begin{equation}
    \frac{\text{d}\rho}{\text{d}t} = \frac{\alpha}{2t} \big(1-\rho^{-1}-\alpha^{-1}\rho\big)\,.
\end{equation}
For $\alpha < 4$, there are no fixed points of the dynamical equation, and every droplet shrinks monotonically and disappears in finite time. 
For $\alpha > 4$, there are two fixed points: an unstable $\rho_1 < \rho_2$ and a stable fixed point $\rho_2$ (note the numerical values of these fixed points are the same as the roots discussed before). 
If we start at $\rho(0) > \rho_1$, the trajectory is pulled into the stable fixed point $\rho_2$. 
The approach towards $\rho_2$ is exponential in time, and since each droplet exponentially locks at $R = R_\text{c}\rho_2$, there is no significant population of droplets that can shrink and dissolve. 
If we have such a fixed point structure with stable and unstable fixed points, the droplet number density cannot decrease. 
This property contradicts our result of a decreasing droplet number density  (Fig.~\ref{fig:em-intlim}(c)). 

For $\alpha = 4$, the two roots merge, and the fixed point is a saddle-node. Due to a non-linear term that now dominates the dynamical equation, the approach towards the root is a slow algebraic relaxation. Since exponential locking is impossible, droplets do not synchronize exponentially fast to a common size, but instead, the growth of some droplets is compensated by the shrinkage of others. This allows for the algebraic decrease of the droplet number density in time.

(iii) For $\Delta = 0$ corresponding $\alpha = 4$, the two roots coincide $\rho_1 = \rho_2 = 2$. The solution is: 
\begin{equation}
\frac{h(\rho)}{\text{C}} = \bigg(\frac{2}{2-\rho}\bigg)^3\, \exp\bigg(\frac{-2}{2-\rho}\bigg)\,.
\label{eq:app:solh}
\end{equation}
Near $\rho \rightarrow 2$, the exponential singularity blows up faster than the power-law singularity. It maintains $h(\rho)\rightarrow 0$ from one side, for $\rho \nearrow 2$ , while $h(\rho)\rightarrow \infty$ from the other. All the derivatives of the solution are zero at $\rho = 2$. 

We conclude that the only physical solution of $h(\rho)$ exists on a finite domain $0\leq\rho < 2$, for the choice of $\alpha = 4$, while $h(\rho) = 0$ for $\rho \geq 2$. Using Eq.~\eqref{eq:app:solh} and the definition of $\alpha = \int_0^2\rho h(\rho) d\rho$, with the choice $\alpha = 4$, we find Eq.~\eqref{eq:int-lim-sol-prev}.

\section{Solution of the coarse-grained background concentration in the quasi-static limit}\label{app:quasi-static-conc}

In the following, we will calculate the coarse-grained background concentration $\bar{c}(t)$ in the quasi-static limit.
For constant matter supply, in the quasi-static limit, we have shown that the rate of change of the droplet phase volume fraction $\text{d}\Phi(t)/\text{d}t$ (Eq.~\eqref{eq:cons-cont-const}), is proportional to the matter supply rate $J/c^\text{in,(0)}$. From the definition of the droplet phase volume fraction $\Phi(t)$ (Eq.~\eqref{eq:C2:def_Phi}), its rate of change is given by
\begin{equation}
\begin{split}
    \frac{\text{d}}{\text{d}t}\Phi(t) &=  {4\pi}\int_0^{R_\text{max}} dR \,   R^{2}(t)\frac{\text{d}R(t)}{\text{d}t}\mathcal{N}(R,t) \\
    &+\frac{4\pi}{3}\int_0^{R_\text{max}} dR\,R^{ 3}(t)\partial_t\mathcal{N}(R,t)\,.
\end{split}    
\end{equation}
From the discussion in Appendix~\ref{app:critical_rad2}, we know that the distribution function lives on a finite support $R\in[ 0, R_\text{max} ]$, where $\mathcal{N}(R,t)$ vanishes at the lower and upper bounds.

Using the continuity equation Eq.~\eqref{eq:em-cont}, we rewrite the partial derivative $\partial_t\mathcal{N}(R,t)$, which leads to:
\begin{equation}
\begin{split}
    \frac{\text{d}}{\text{d}t}\Phi(t) &=  {4\pi}\int_0^{R_\text{max}} dR \,  R^{2}(t)\frac{\text{d}R(t)}{\text{d}t}\mathcal{N}(R,t) \\
    &-\frac{4\pi}{3}\int_0^{R_\text{max}} dR\,R^{3}(t)\partial_{R}\bigg(\frac{\text{d}R(t)}{\text{d}t}\mathcal{N}(R,t)\bigg)\,.
\end{split}    
\end{equation}
Performing partial integration of the second term in the sum gives:
\begin{equation}
\begin{split}
    \frac{\text{d}}{\text{d}t}\Phi(t) &=   {8\pi}\int_0^{R_\text{max}} dR \, R^{2}(t)\frac{\text{d}R(t)}{\text{d}t}\mathcal{N}(R,t)
    \\
    &-\left.\frac{4\pi}{3}R^{3}(t)\frac{\text{d}R(t)}{\text{d}t}\mathcal{N}(R,t)\right|_0^{R_\text{max}}\,.
\end{split}
\label{eq:app-dphi}
\end{equation}
 For $R=0$, there are no more droplets due to the dissolution, such that $\mathcal{N}(R=0,t)=0$. For small radii, the growth law $\text{d}R/\text{d}t \propto - R^{-1}(1+\beta(R))^{-1}$. Thus, the term $R^3(t)\text{d}R/\text{d}t \propto R, R^2$, in the diffusion-limited and interface-resistance-limited regime, respectively, and it vanishes for $R=0$. For the upper limit, we consider a finite cut-off $R_\text{max}$, at which the droplet size density vanishes. Furthermore, the terms $R^3(t)\text{d}R/\text{d}t \propto R^2$ and $R^3$ in the diffusion-limited and interface-resistance-limited regimes, respectively, are well-defined for large and finite droplets. Thus, the boundary term in Eq.~\eqref{eq:app-dphi} vanishes. 

We are only interested in the interface-resistance-limited regime. Using for the r.h.s. of Eq.~\eqref{eq:cons-cont-const}, the rate of change of the droplet phase volume fraction $\text{d}\Phi/\text{d}t$ (Eq.~\eqref{eq:app-dphi}), and the growth law $\text{d}R/\text{d}t$ (Eq.~\eqref{eq-drdt00} for $\beta \ll 1$), we solve for the coarse-grained background concentration $\bar{c}(t)$ in the quasi-static limit:
\begin{equation}
\frac{\bar{c}(t)}{c^{(0)}} = 1 + \ell_\gamma \frac{\langle R(t)\rangle}{\langle R(t)^2 \rangle}+\frac{ J\big( 8\pi \, k c^{(0)}\big)^{-1}}{\langle R(t)^2 \rangle n(t) }\,,
\label{eq:barc-Jc-app}
\end{equation}
where we have used the definition of the $k$-th moment Eq.~\eqref{eqC2:kth_moment}  and the number density Eq.~\eqref{eqC2:number_cont}.

The background concentration $\bar{c}(t)$ is set by two contributions, one related to the droplet size density $\mathcal{N}(R,t)$ and the other one arising from the matter supply density $J$. Similarly, both contributions affect the critical radius $R_\text{c}(t) = \ell_\gamma c^{(0)}/(\bar{c}(t)-c^{(0)})$. Without matter supply ($J=0$), we obtain the solution of Wagner~\cite{wagner_theorie_1961} for the quasi-static solution of the coarse-grained background concentration in a passive emulsion with a conserved droplet phase volume fraction $\Phi(t)$.

Inserting the solution Eq.~\eqref{eq:barc-Jc-app} into Eq.~\eqref{eq-drdt00}, and using the Gibbs-Thomson relation for the equilibrium concentration $c^\text{eq}(R)$ (Eq.~\eqref{eq:ceq_finite}), gives Eq.~\eqref{eq:dR-full}. It describes the droplet growth in the quasi-static limit for $\beta \ll 1$ and $J=\text{const}$.

\section{Parameter values}

\begin{table*}[!t]
\centering
\caption{\textbf{Parameter choices for single droplet dynamics.} Time and radius are nondimensionalized as $t/\tau$, and $R/\ell_\gamma$. For the nondimensional matter supply density $J/J_0$, where $J_0 = c^{(0)}/\tau$.}
\label{tab-param1}
\setlength{\tabcolsep}{3.5pt}
\begin{tabular}{l c c c c c c c c c c}
\hline 
Figure & $V_\text{d}(0)/V_\text{sys}$ & $c^\text{(0)}/c^\text{in,(0)}$ & $\varepsilon(0)$ & $R_\text{c}(0)/\ell_\gamma$ & $J/J_0$ & Comments \\
\hline
Fig.~\ref{fig2-diff-lim}(d,e) & $1.75\cdot 10^{-5}$ & $10^{-3}$ & $0.07$ & $14.28$ & $2.82$ & $\tau = \ell_\gamma^2/D^\text{out}$ & \\[0.7em]
Fig.~\ref{int-lim1} & $1.75\cdot 10^{-5}$ & $10^{-3}$ & $0.07$ & $14.28$ & $0.564$ & $\tau = \ell_\gamma/k$ 
\\
\hline
\end{tabular}
\end{table*}

\begin{table*}[!t]
\centering
\caption{\textbf{Parameter choices for the dynamics of two droplets.} Time and radius are nondimensionalized as $t/\tau$, where $\tau = \ell_\gamma^2/D^\text{out}$, and $R/\ell_\gamma$.}
\label{tab-param2}
\setlength{\tabcolsep}{3.5pt}
\begin{tabular}{l c c c c c c c c c c}
\hline 
Figure & $V_\text{sys}/\ell_\gamma^3$ & $c^\text{(0)}/c^\text{in,(0)}$ & $\varepsilon(0)$ & $R_\text{c}(0)/\ell_\gamma$ & Comments  \\
\hline
Fig.~\ref{fig3-diff-lim-1} & $2.68\cdot10^8$ & $2.5\cdot 10^{-3}$ & $0.16$ & $6.07$ & (c,d) $Jt/c^{(0)} = 0.064$
\\
\hline
\end{tabular}
\end{table*}

\begin{table*}[!t]
\centering
\caption{\textbf{Parameter choices for the dynamics of emulsions.} Time and radius are nondimensionalized as $t/\tau$, where $\tau = \ell_\gamma^2/D^\text{out}$, and $R/\ell_\gamma$, where $D^\text{out}$ and $\ell_\gamma$ are fixed. For the nondimensional matter supply density $J/J_0$, where $J_0 = D^\text{out}c^{(0)}/\ell_\gamma^2$. The initial droplet radii are initialized by the choice of $\mathcal{N}(R,0) R_\text{c}(0)/n(0)$ either Eq.~\eqref{eq:passive-distr-diff} or Eq.~\eqref{eq:passive-distr-int} depending if $k R_\text{c}(0)/D^\text{out} > 1$ or $k R_\text{c}(0)/D^\text{out} < 1$ respectively.}
\label{tab-param}
\setlength{\tabcolsep}{3.5pt}
\begin{tabular}{l c c c c c c c c c}
\hline 
Figure & ${k \ell_\gamma}/{D^\text{out}}$ & $c^\text{(0)}/c^\text{in,(0)}$ & $R_\text{c}(0)/\ell_\gamma$ & $\langle R(0)\rangle/\ell_\gamma$  &Comments \\
\hline
Fig.~\ref{fig:const-eps} & $[ 10^{-6},10^{-2},1, 10^{4}]$ & $5\cdot 10^{-3}$  & $0.33$ & $0.33$ & varying ${k}$ \\[0.7em]
Figs.~\ref{fig:em-diff-lim},~\ref{fig:C2_Rc_const}(a) & $10^{12}$ & $5\cdot 10^{-3}$  & $0.33$ & $0.33$ & varying $J/J_0$\\[0.7em]
Figs.~\ref{fig:em-intlim},~\ref{fig:C2_Rc_const}(b) & $10^{-3}$ & $5\cdot 10^{-3}$ & $0.33$ & $0.29$ & varying $J/J_0$ 
\\
\hline
\end{tabular}
\end{table*}


\end{document}